\documentclass[10pt,aps,prl,onecolumn,superscriptaddress,floatfix,nofootinbib,notitlepage,showpacs,preprintnumbers]{revtex4-1}
\usepackage{amsmath,amssymb,amsfonts}
\usepackage{graphicx,color}

\newcommand{\avg}[1]{\langle #1\rangle}

\begin{document}

\title{Statistically validated network of portfolio overlaps and systemic risk}

\author{Stanislao Gualdi}
\affiliation{\small Laboratoire de Math\'ematiques Appliqu\'ees aux Syst\`emes, Centrale Sup\'elec, 92290 Ch\^atenay-Malabry, France}
\author{Giulio Cimini}
\affiliation{\small IMT School for Advanced Studies, 55100 Lucca, Italy}
\affiliation{\small Istituto dei Sistemi Complessi (ISC)-CNR, 00185 Rome, Italy}
\author{Kevin Primicerio}
\affiliation{\small Laboratoire de Math\'ematiques Appliqu\'ees aux Syst\`emes, Centrale Sup\'elec, 92290 Ch\^atenay-Malabry, France}
\author{Riccardo Di Clemente}
\affiliation{\small MIT - Massachusetts Institute of Technology, Cambridge, MA 02139, United States}
\author{Damien Challet}
\affiliation{\small Laboratoire de Math\'ematiques Appliqu\'ees aux Syst\`emes, Centrale Sup\'elec, 92290 Ch\^atenay-Malabry, France}
\affiliation{\small Encelade Capital SA, 1015 Lausanne, Switzerland}

\begin{abstract}
Common asset holding by financial institutions, namely {\em portfolio overlap}, is nowadays regarded as an important channel for financial contagion 
with the potential to trigger fire sales and thus severe losses at the systemic level. In this paper we propose a method to assess the statistical significance 
of the overlap between pairs of heterogeneously diversified portfolios, which then allows us to build a validated network of 
financial institutions where links indicate potential contagion channels due to realized portfolio overlaps. 
The method is implemented on a historical database of institutional holdings 
ranging from 1999 to the end of 2013, but can be in general applied to any bipartite network where the presence of similar sets of neighbors is of interest.
We find that the proportion of validated network links ({\em i.e.}, of statistically significant overlaps) increased steadily before the 2007-2008 global financial crisis and reached a maximum 
when the crisis occurred. We argue that the nature of this measure implies that systemic risk from fire sales liquidation was maximal at that time. After a sharp drop in 2008, 
systemic risk resumed its growth in 2009, with a notable acceleration in 2013, reaching levels not seen since 2007. 
We finally show that market trends tend to be amplified in the portfolios identified by the algorithm, such that it is possible to have an informative signal about 
financial institutions that are about to suffer (enjoy) the most significant losses (gains). 
\end{abstract}

\maketitle

\section*{Introduction}\label{sec:intro}

The 2007-2008 global financial crisis has drawn the attention of both academics and regulators to the complex interconnections between financial institutions~\cite{glasserman2015contagion} 
and called for a better understanding of financial markets especially from the viewpoint of systemic risk, {\em i.e.}, the possibility that a \emph{local} event triggers a \emph{global} instability through a cascading 
effect~\cite{Lau2009IMF,Brunnermeier2009JEP,Gai2010PRSA,Staum2013hsr,Acemoglu2015AER,Battiston2016Science}.
In this respect, while much effort has been devoted to the study of counter-party and roll-over risks caused by loans between 
institutions~\cite{Allen2000JPE,Eisenberg2001MS,Iori2006JEBO,May2010JRSI,Haldane2011Nature,Bluhm2011CFS,Krause2012JEBO,Cimini2015SR,Cimini2016arXiv,Barucca2016arXiv}, 
the ownership structure of financial assets has received relatively less attention, primarily because of lack of data and of adequate analysis techniques.
Yet, while in traditional asset pricing theory assets ownership does not play any role, there is increasing evidence that it is a potential source of non-fundamental risk and, 
as such, can be used for instance to forecast stock price fluctuations unrelated to fundamentals~\cite{Greenwood2011JFE,Anton2014JF}. 
More worryingly, if investment portfolios of financial institutions are too similar (as measured by the fraction of common asset holdings, or portfolio overlap), 
the unexpected occurrence of financial distress at the local level may trigger fire sales, namely assets sales at heavily discounted prices. 
Fire sales spillovers are believed to be an important channel of financial contagion contributing to systemic risk~\cite{Shleifer1992JF,Cifuentes2005JEEA,Shleifer2010JEP,Caccioli2014JBF,Cont2014MF,Greenwood2015JFE}: 
when assets prices are falling, losses by financial institutions with overlapping holdings become self-reinforcing and trigger further simultaneous sell orders, 
ultimately leading to downward spirals for asset prices. From this point of view, even if optimal portfolio selection helps individual firms to diversify risk, it can also 
make the system as a whole more vulnerable~\cite{Corsi2013,glasserman2015contagion}.
The point is that fire sale risk builds up gradually but reveals itself rapidly, generating a potentially disruptive market behavior.

In this contribution we propose a new statistical method to quantitatively assess the significance of the overlap between a pair of portfolios, with the 
aim of identifying those overlaps bearing the highest riskiness for fire sales liquidation. Since we apply the method to institutional portfolios we will use 
interchangeably the terms institution and portfolio throughout the paper. 
In practical terms, the problem consists in using assets ownership data by financial institutions to establish links between portfolios having strikingly similar pattern of holdings. 
Market ownership data at a given time $t$ consists of a set $I(t)$ of institutions, holding positions from a universe of $S(t)$ securities (or financial assets in general). 
The $|I(t)|\times |S(t)|$ ownership matrix $\mathcal{W}(t)$ describes portfolios composition: its generic element $W_{is}(t)$ denotes the number of shares of security $s\in S(t)$ held by institution $i\in I(t)$. 
The matrix $\mathcal{W}(t)$ can be mapped into a binary ownership matrix $\mathcal{A}(t)$, whose generic element $A_{is}(t)=1$ if $W_{is}(t)>0$ and $0$ otherwise, 
which allows to define the degree $d_i(t) = \sum_s A_{is}(t)$ of an institution $i$ as the number of securities it owns at time $t$, 
and the degree $d_s(t) = \sum_i A_{is}(t)$ of a security $s$ is the number of investors holding it at time $t$. The number of securities held by both institutions $i$ and $j$, 
namely the {\em overlap} of their portfolios, is instead given by $o_{ij}(t)=\sum_sA_{is}(t)A_{js}(t)$ (with $i\neq j$), which is the generic element of the $|I(t)|\times |I(t)|$ portfolio overlap matrix $\mathcal{O}(t)$.  
In network theory language, $\mathcal{O}(t)$ represents a {\em projected} monopartite network of institutions obtained as a contraction of the binary ownership matrix $\mathcal{A}(t)$, 
which instead represents a bipartite network of institutions and securities. However, in such a projected network two institutions are connected as soon as they invest in the same security: 
this generates too many links and fails to filter out less risky overlaps. For example, a security held by a large number of investors would trivially determine a correspondent number of projected links 
without a clear meaning. Although there is no unique way to tackle this problem, the point of view we take here can be roughly summarized as follows: 
if we were to reshuffle links in the original bipartite network without changing the degree of each node, how likely is the observed overlap? 
Thus, the problem is that of building a {\em validated} projection of the original bipartite network containing only the most {\em significant} overlaps 
that cannot be explained by a proper null network model. In this way we can drastically reduce the original amount of links and obtain a much sparser validated network with a clearer meaning.

All methods to build validated projections proposed in the literature involve the use of a threshold to determine which links are retained in the monopartite network, 
but vary in how the threshold is chosen~\cite{Neal2014SN}. The simplest and most common approach is to use an unconditional global threshold~\cite{Latapy2008SN,Neal2013SNAM}, 
which however suffers from arbitrariness, structural bias and uniscalarity---by systematically giving preference to institutions with many holdings~\cite{Neal2014SN}. 
Using a threshold which depends on institution degrees can overcome the latest two limitations~\cite{Serrano2009PNAS,Borgatti2011SAGE}. In particular, 
the threshold can be determined using a null hypothesis of random institutions-to-securities matching constrained to institutions degree, 
for which the probability that two institutions share a given number of securities is given by a hypergeometric distribution~\cite{Sudarsanam2002GR,Goldberg2003PNAS}. 
Yet, also this approach is biased by implicitly treating securities as equivalent and interchangeable. A recently proposed improvement to this method consists in building homogeneous networks of securities, 
that is, in splitting the original bipartite network into subnetworks each consisting of securities with the same degree and of all institutions linked to them~\cite{Tumminello2011PLoS}. 
In this way, the null hypothesis can be properly cast, for each layer separately, with the hypergeometric distribution. 
Problems however arise when securities are characterized by a strongly heterogeneous number of investors: the process of creating homogeneous subnetworks 
with securities having the same degree often translates into almost empty subsets, causing a serious resolution problem and leading to almost empty validated networks (see section Methods). 
A possible solution here is to perform link validation without taking into account degree heterogeneity~\cite{Tumminello2011PLoS}, 
which however cannot be formalized analytically since the events of choosing different securities have now different occurrence probabilities. 
An alternative approach consists in using a null model of random institutions-to-securities matching constrained not only to institutions degree but also to securities degree. 
The fixed degree sequence model (FDSM)~\cite{Zweig2011SNAM,Horvat20131SNAM} and the stochastic degree sequence model (SDSM)~\cite{Neal2014SN} belong to this category. 
In the FDSM, the null hypothesis cannot be formalized analytically and the method relies on a conditional uniform graph test 
by generating a microcanonical ensemble of random graphs whose overlaps can be compared with the empirical ones. 
However, algorithms to sample the graph configuration space are impractically complex or biased~\cite{Blanchet2013RSA}, or suffer from arbitrariness~\cite{Gionis2007ACM}. 
In contrast, in the SDSM  the null hypothesis can be formalized at least numerically, but it is computationally impractical in most cases~\cite{Neal2014SN}. 
Thus, also the SDSM  relies on a conditional uniform graph test, which is however easy to achieve by using the linking probabilities between institutions and securities 
obtained with the Link Probability Model (LPM)~\cite{mcCulloh2010SSRN}. Yet, in the LPM these probabilities are basically the proportion of link occurrences over multiple observations of the data, 
which requires much more information than that contained in the ownership matrix, and, more importantly, represents a valid approach only when the underlying network is assumed to be stationary 
in time---which is clearly not the case for stock markets.

The method we propose here overcomes all the limitations of its predecessors by building on a null hypothesis described by the {\em Bipartite Configuration Model} 
(BiCM)~\cite{Saracco2015SR}, which is the extension of the standard {\em Configuration Model} \cite{PhysRevE.70.066117} to bipartite graphs. 
In the null BiCM network, institutions randomly connect to securities, but the degrees of both institutions and securities are constrained on average to their observed values in real ownership data. 
This is achieved through maximization of the Shannon entropy of the network subject to these constraints, which remarkably allows to analytically and numerically formalize the null hypothesis (see section Methods). 
The additional advantages of the BiCM with respect to~\cite{Tumminello2011PLoS} is that of not requiring the homogeneity of neither layer of the network, 
and with respect to~\cite{Neal2014SN,mcCulloh2010SSRN} of using only the information contained in a single snapshot of the data. The method works as follows. For each date $t$, 
in order to distinguish the true signal of overlapping portfolios from the underlying random noise, every link of the projected network has to be independently validated 
against the BiCM null hypothesis. Thus, for each pair of institutions $(i,j)$ having overlap $o_{ij}(t)$, we compute the probability distribution $\pi(\cdot|i,j,t)$ of the expected overlap 
under the BiCM (see section Methods). The statistical significance of $o_{ij}(t)$ is then quantified through a p-value:
\begin{equation}
P[o_{ij}(t)] = 1-\sum_{x=0}^{o_{ij}(t)-1}\pi(x|i,j,t),
\label{eq.pval}
\end{equation}
where the right-hand size of Eq.~(\ref{eq.pval}) is the cumulative distribution function of $\pi(\cdot|i,j,t)$, namely the probability to have an overlap larger or equal than the observed one under the null hypothesis. 
If such a p-value is smaller than a threshold $P^*(t)$ corrected for multiple hypothesis testing (see section Methods), we validate the link between $i$ and $j$ and place it on the monopartite validated network of institutions.
Otherwise, the link is discarded. In other words, the comparison is deemed statistically significant if the observed overlap would be an unlikely realization of the null hypothesis according to the significance level $P^*(t)$.
This procedure is repeated for all pairs of institutions, resulting in the validated projection $\mathcal{V}(t)$ of the original network: a monopartite network whose generic element 
$V_{ij}(t)=1$ if $P[o_{ij}(t)]<P^*(t)$, and $0$ otherwise.

\begin{figure}[t!]
\centering
 \includegraphics[scale=0.25]{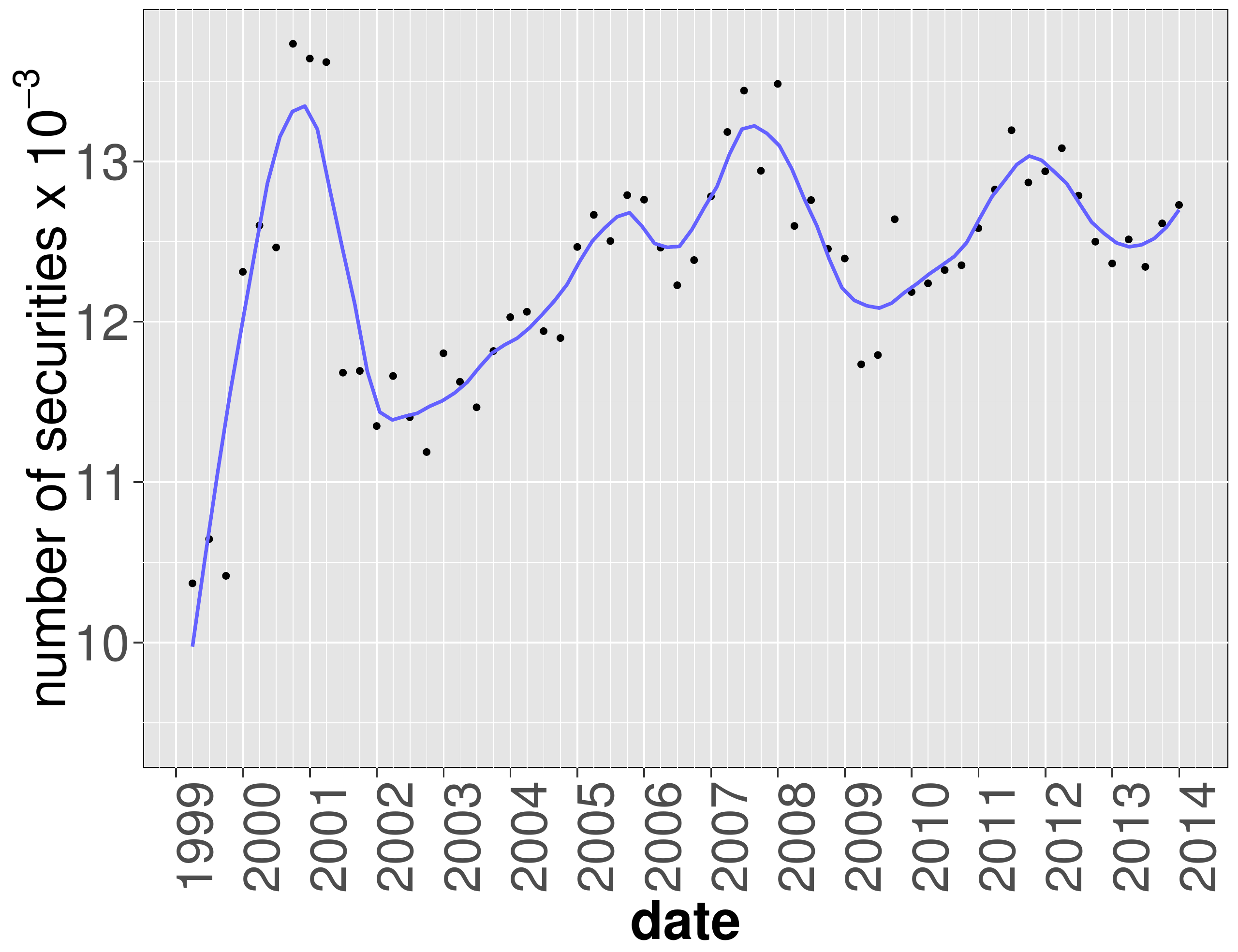}\quad
 \includegraphics[scale=0.25]{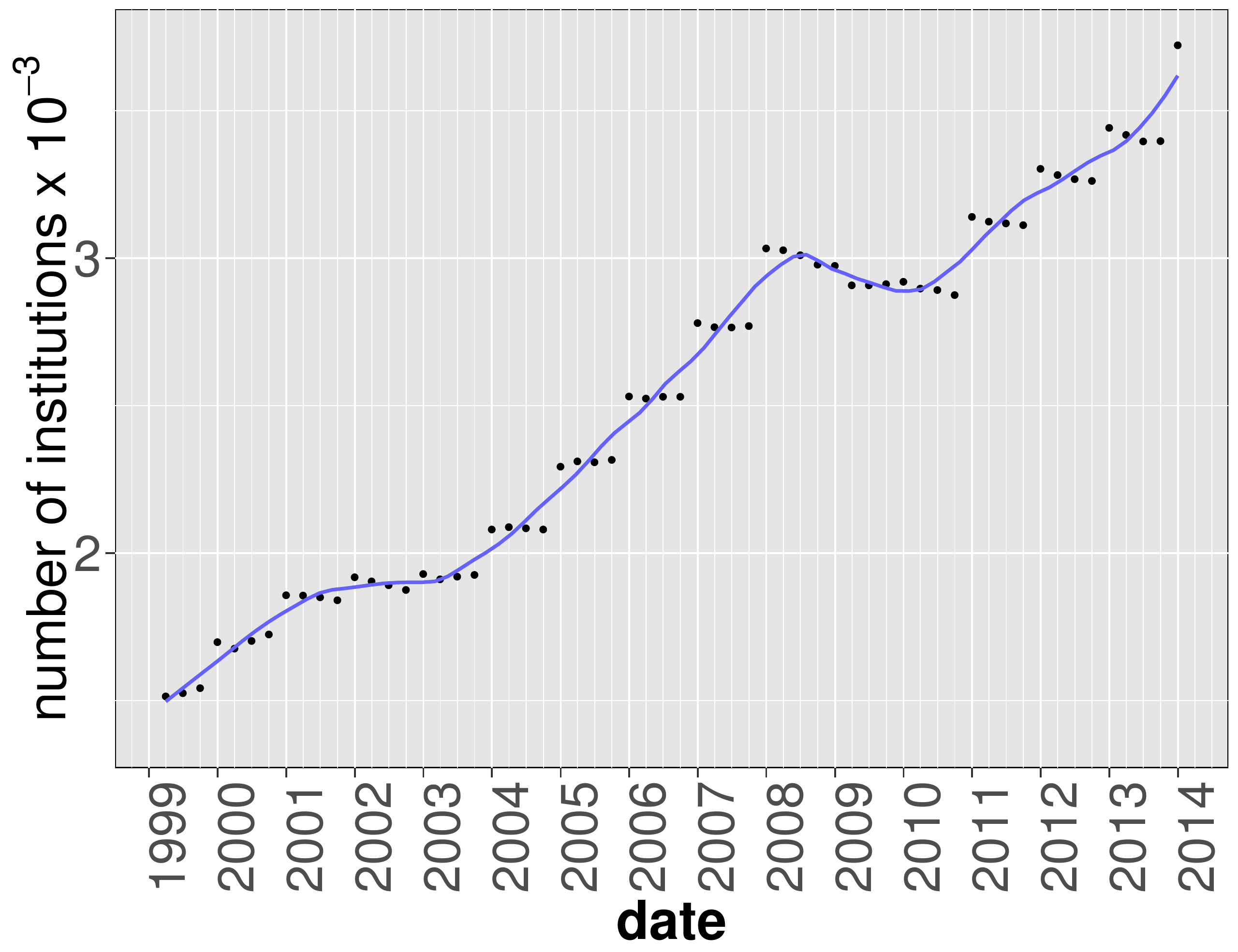}\\
 \includegraphics[scale=0.25]{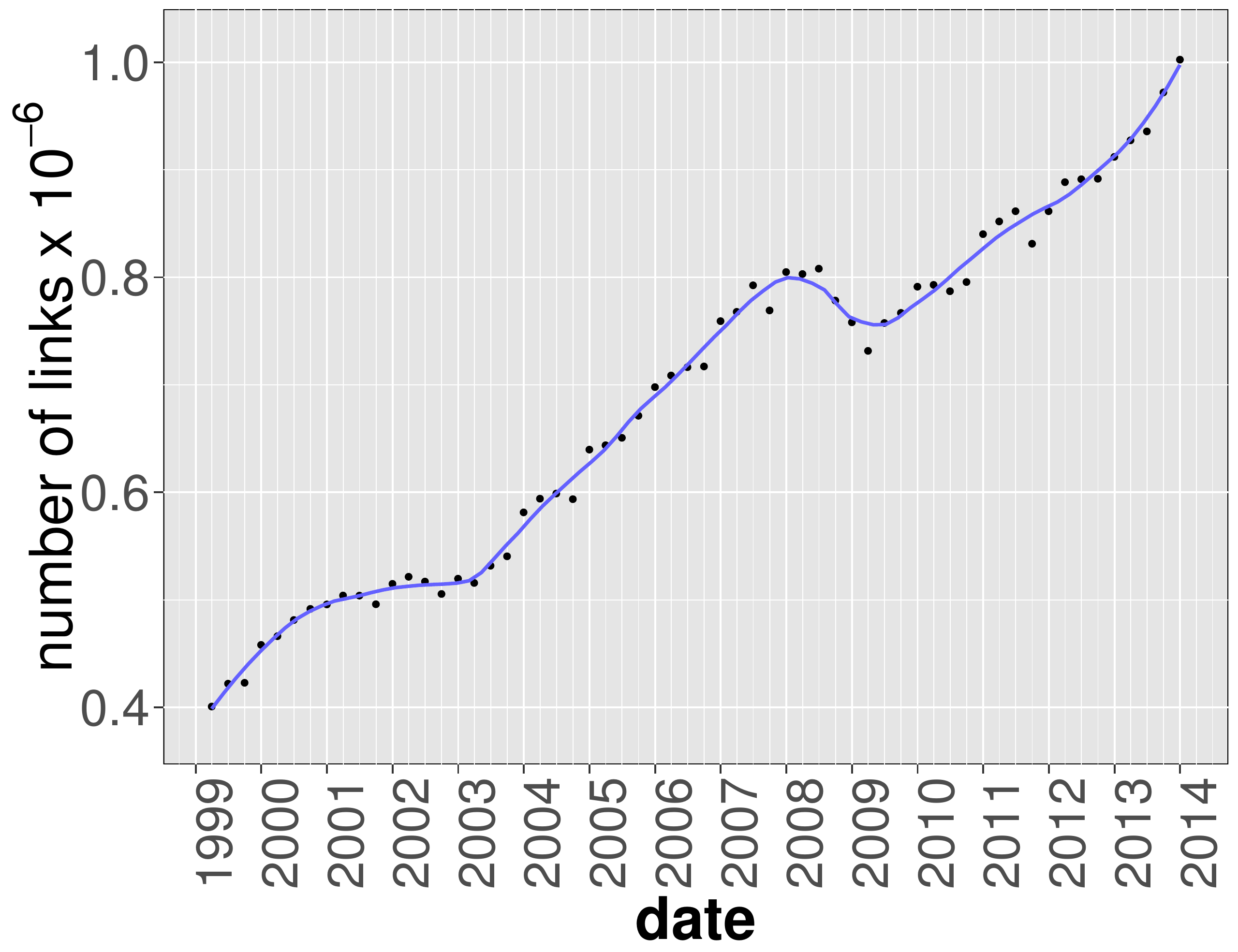}\quad
 \includegraphics[scale=0.25]{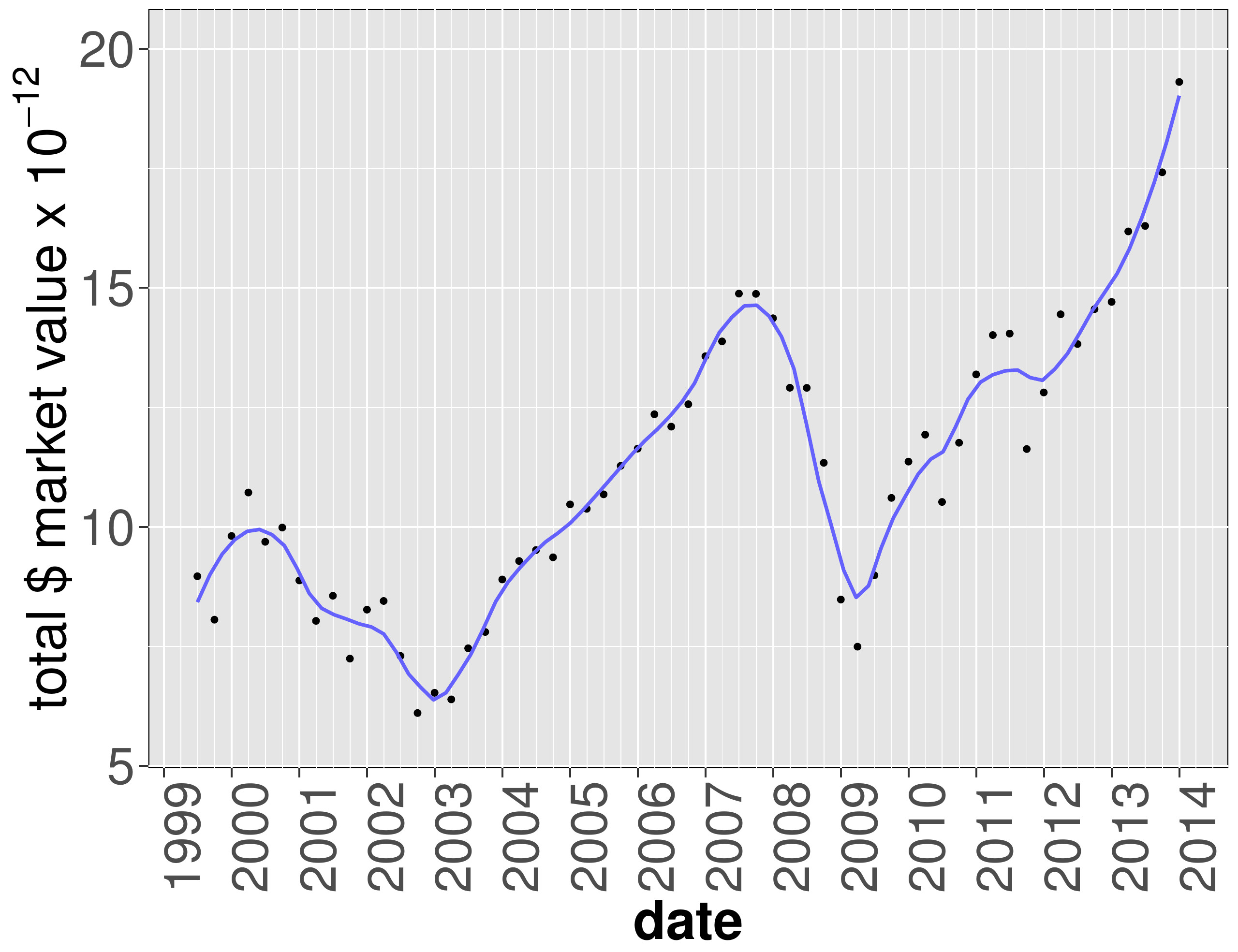}
\caption{Temporal evolution of main aggregate quantities characterizing the bipartite ownership network. From left to right: 
number of institutions $|I(t)|$, number of securities $|S(t)|$, number of different ownership relations $L(t)=\sum_{is}A_{is}(t)$ and total market value $MV(t)=\sum_{is}W_{is}(t)p_s(t)$, 
where $p_s(t)$ and $\sigma_s(t)=\sum_{i}W_{is}(t)$ are the price and number of outstanding shares of security $s$ at time $t$, respectively. Solid lines correspond to a locally weighted least squares regression 
(loess) of data points with 0.2 span.}
\label{fig:general_info}
\end{figure}

When applied to a historical database of SEC 13-F filings (see section Methods for details and Fig.~\ref{fig:general_info} for the temporal evolution of the main dataset statistics), 
our method yields statistically validated networks of overlapping portfolios whose properties turn out to be related to the occurrence of the 2007-2008 global financial crisis. 
In particular, we propose to regard the average number of validated links for each institution as a simple measure of systemic risk due to overlapping portfolios. 
Such a measure gradually built up in years from 2004 to 2008, and quickly dropped after the crisis. Systemic risk has then been increasing since 2009, 
and at the end of 2013 reached a value not previously seen since 2007. Note that because there is only one large crisis in our dataset, 
we refrain from making strong claims about the systematic coincidence of highly connected validated networks and the occurrence of financial crises. 
We also find that overlapping securities ({\em i.e.}, those securities making up the validated overlaps) represent a larger average share of institutional portfolios, 
a configuration which would exacerbate the effect of fire sales. Additionally, we show that the presence of a validated link between two institutions 
is a good indicator of portfolio losses for these institutions in times of bearish markets, and of portfolio growth in times of bullish markets: 
validated links should indeed represent self-reinforcing channels for the propagation of financial distress or euphoria.
More in general, we find that market trends tend to be amplified in the portfolios identified by the algorithm.
Finally, we apply the validation procedure to the overlapping ownerships of securities to identify contagion channels between securities themselves, 
and observe a stable growth of validated securities over the considered time span. This signals an ongoing, deep structural change of the financial market and, 
more importantly, that there are more and more stocks that can be involved in a potential fire sale. 
The presence of local maxima within this trend correspond to all periods of financial turmoil covered by the database: 
the {\em dot-com} bubble of 2001, the 2007-2008 global financial crisis and the 2010-2011 European sovereign debt crisis. 

\section*{Results and Discussion}\label{sec:res}


\begin{figure}[t!]
 \centering
 \vspace{-1cm}
 \includegraphics[scale=0.4]{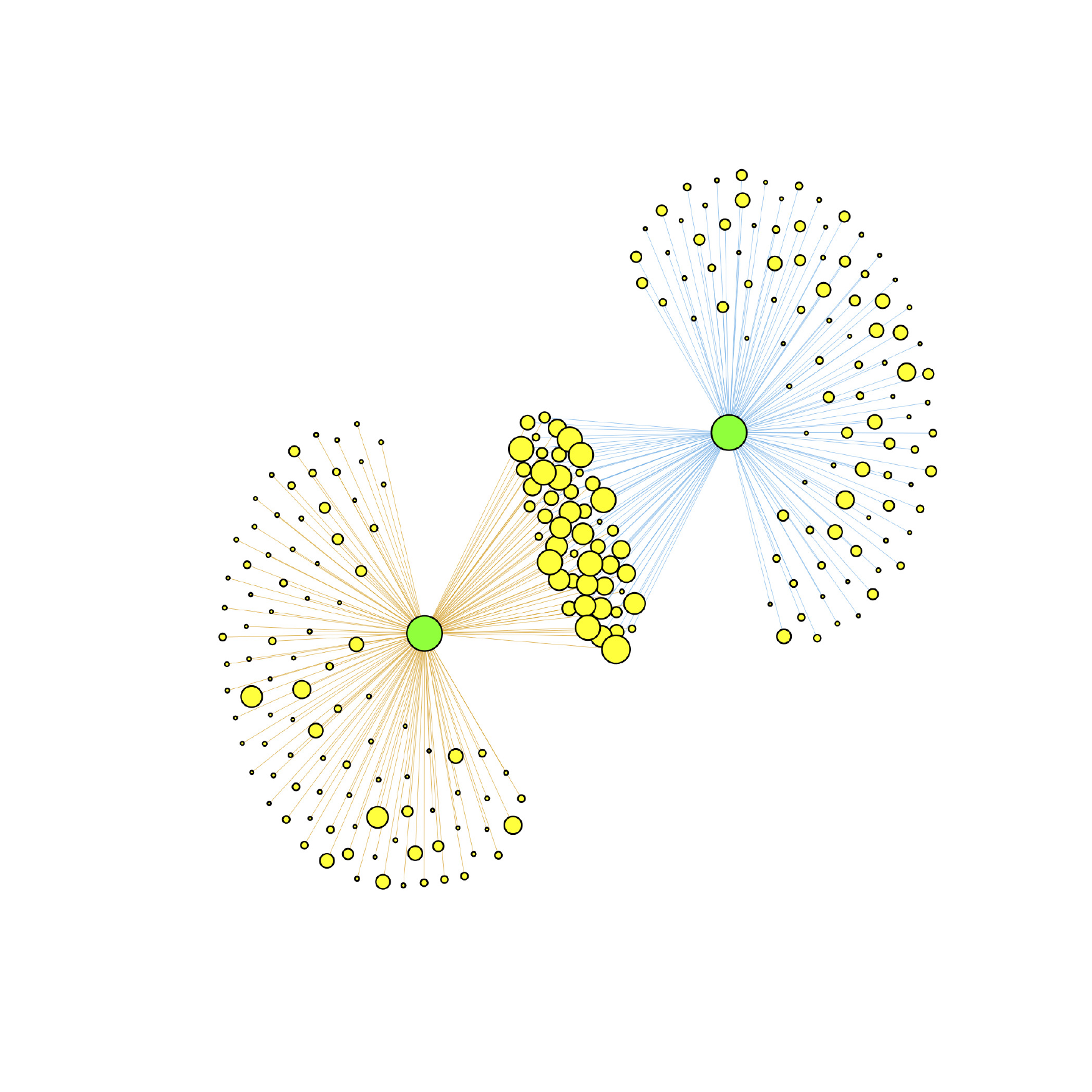}
 \includegraphics[scale=0.4]{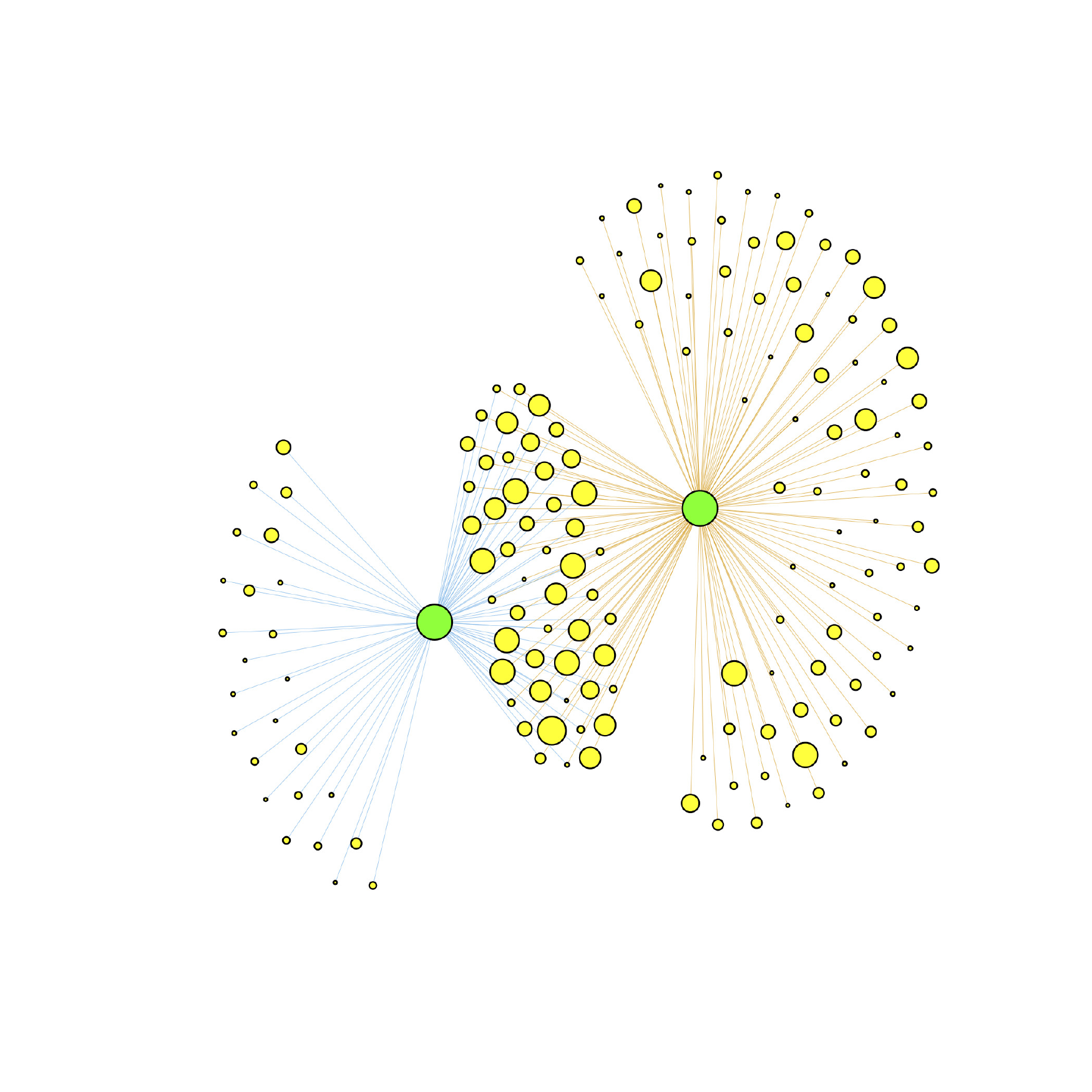}\\
 \vspace{-1cm}
\caption{Two examples of institutions pairs (green nodes) with the securities they own (yellow nodes). 
The composition of each portfolio is denoted by different colors (blue and orange links). The symbol size of a security is proportional to the total number of its investors. 
Although both pairs in the plot have an overlap of 50 securities, the right pair is validated by the algorithm whereas the left pair is not.
This is due to the fact that both the blue and orange portfolios on the left are smaller (the blue one in particular) 
and therefore under the BiCM null model the chance of having the same overlap of the pair on the right is considerably smaller.}
\label{fig:micro}
\end{figure}

\begin{figure}[t!]
\centering
\includegraphics[scale=0.4]{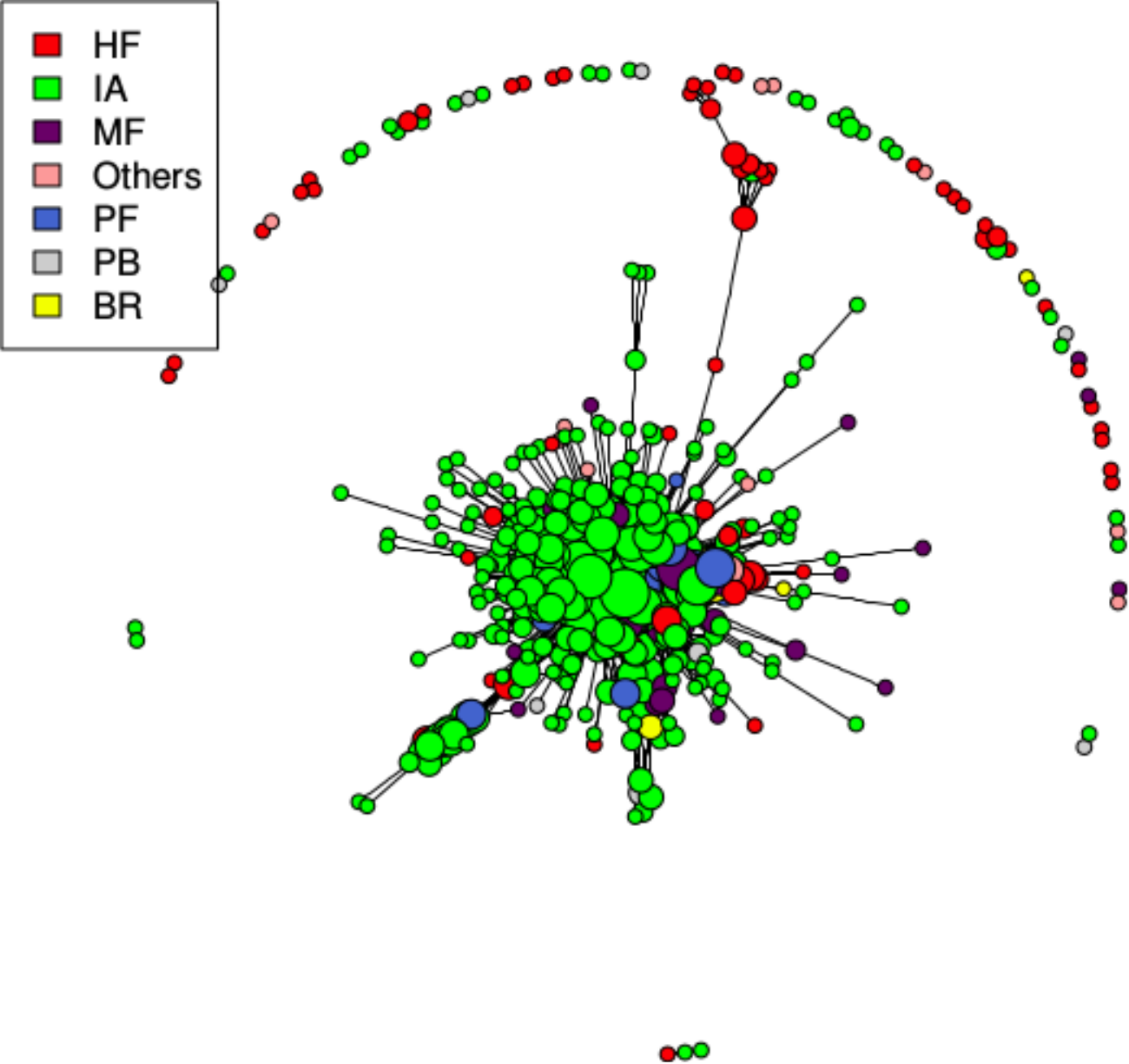}
\caption{Validated networks of institutions at 2006Q4 (1293 institutions and 93602 validated links). 
Node colors label institution type, while their size is given by the logarithm of their degrees in the validated network: $d_i^V(t)=\sum_j V_{ij}(t)$. 
An institution is classified either as Broker (BR), Hedge Fund (HF), Investment Adviser (IA), Mutual Fund (MF), Pension Fund (PF), Private Banking (PB), 
or Other ({\em i.e.}, without classification or belonging to a minor category).}
\label{fig:validated_fancy}
\end{figure}

In order to properly understand the results of our validation method for overlapping portfolios, it is useful to provide a specific example.
Fig.~\ref{fig:micro} shows two similar situations: two pairs of portfolios both owning 50 securities in common. 
Only the right pair is validated by our method, whereas the left pair is not. This happens because the portfolios in the right pair are of smaller size (especially the blue one)
and the same overlap is therefore less likely to happen by chance. Hence, although the algorithm cannot directly take into account how much each institution is investing 
(particularly with respect to the total asset managed by the institution), it does so indirectly by taking into account the diversification of different portfolios ({\em i.e.}, the degree of institutions). 
Validated pairs of portfolios indeed correspond to overlaps which constitute a considerable fraction of the total portfolio value of the pair. 
In short, pairs are validated when neither the diversification of the investments nor the degree of the securities are sufficient to explain the observed overlap.
As we shall see later, the same mechanism is at play when we project the bipartite network on the securities side.
In this case, since the degree of a security is a good approximation of its capitalization and of the dilution of its ownership \cite{Zumbach2004QF,Eisler2006EPJB},
the method will tend to validated links among securities whose ownership is relatively concentrated.
Fig.~\ref{fig:validated_fancy} gives an overall picture of how the validated network looks like.
In general, we observe the presence of multiple small clusters of institutions, together with a significantly larger cluster composed by many institutions linked by a complex pattern of significant overlaps.

\subsection*{Temporal evolution of the validated network of institutions}

\begin{figure}[t!]
\centering
 \includegraphics[scale=0.25]{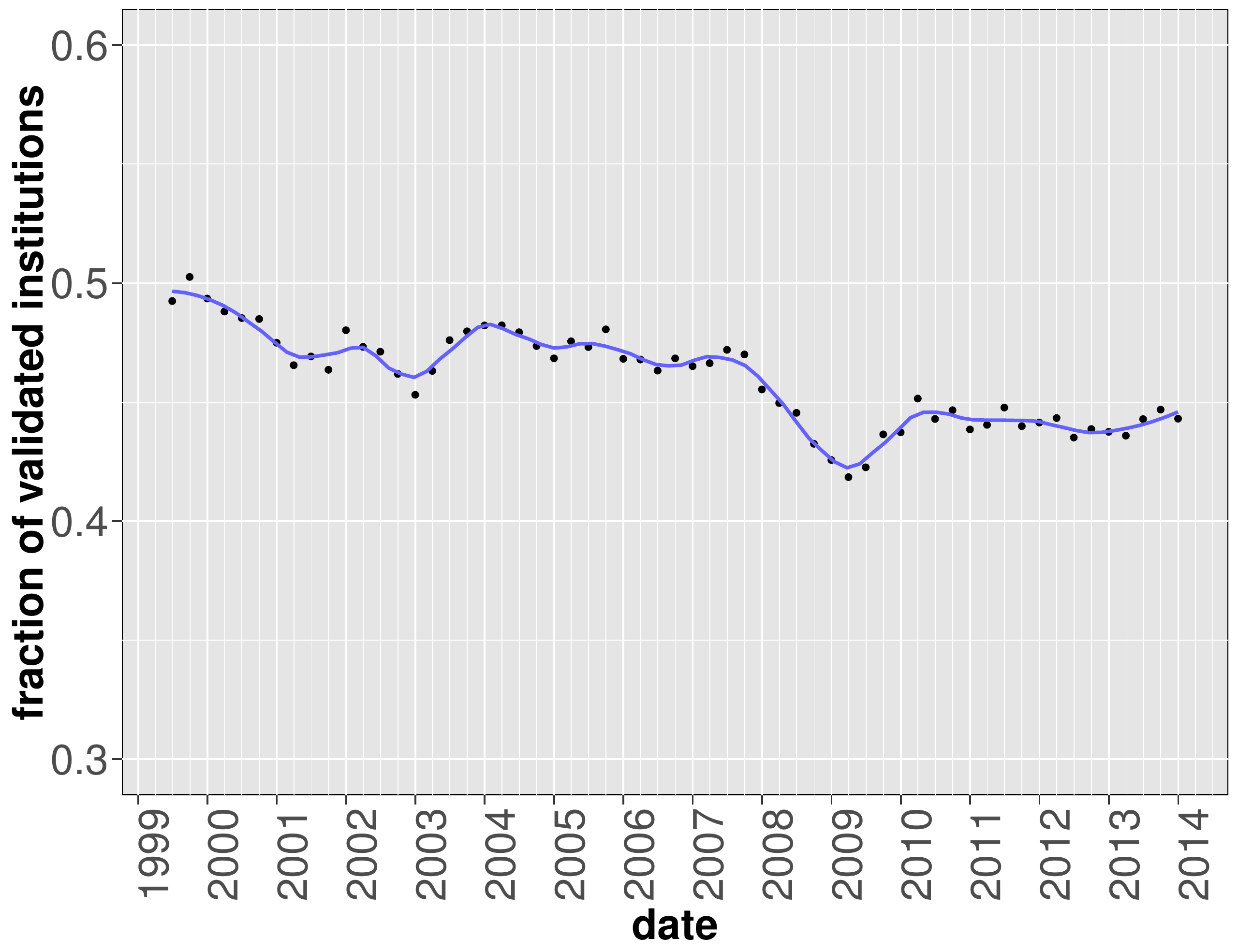}\\
 \includegraphics[scale=0.25]{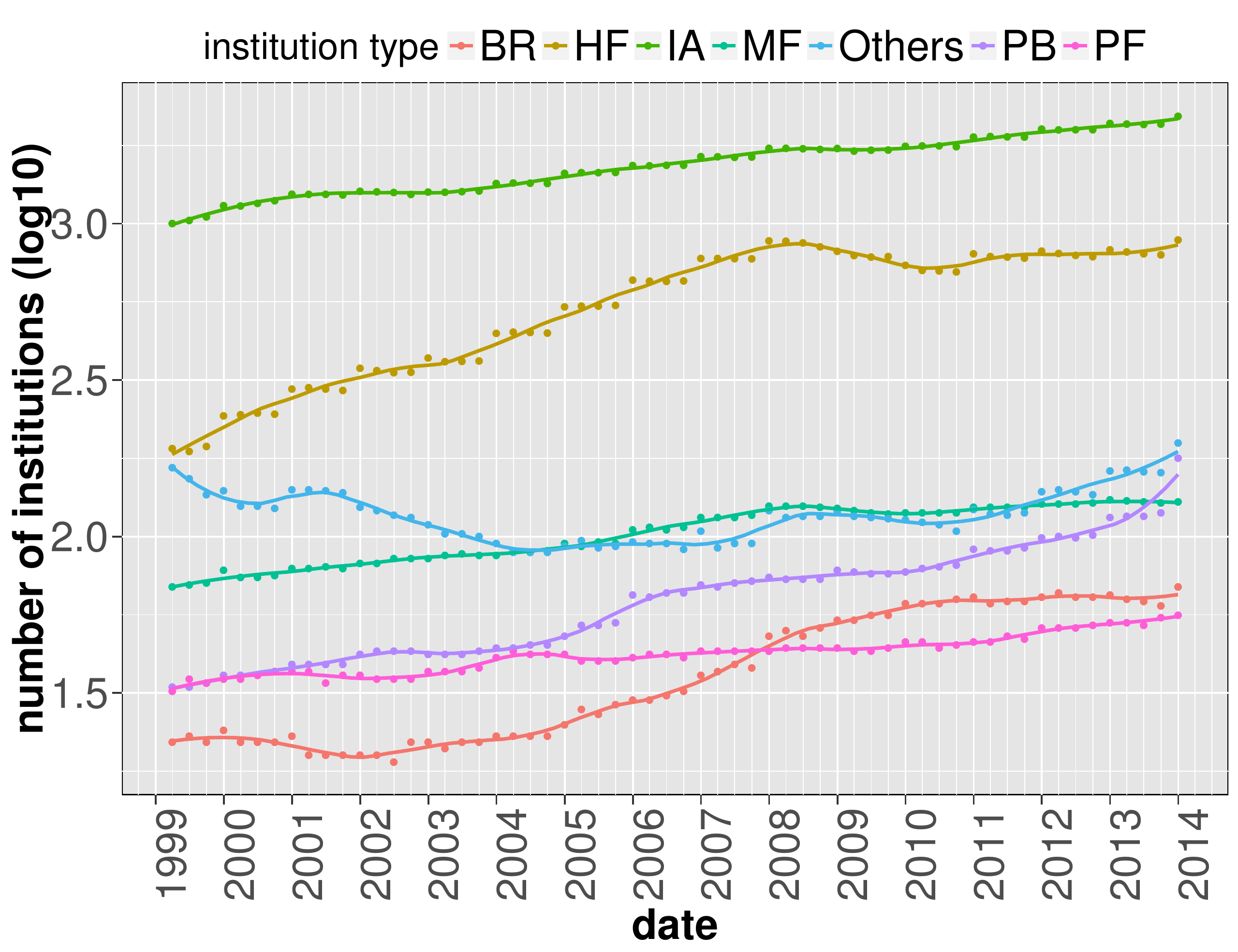}\quad
 \includegraphics[scale=0.25]{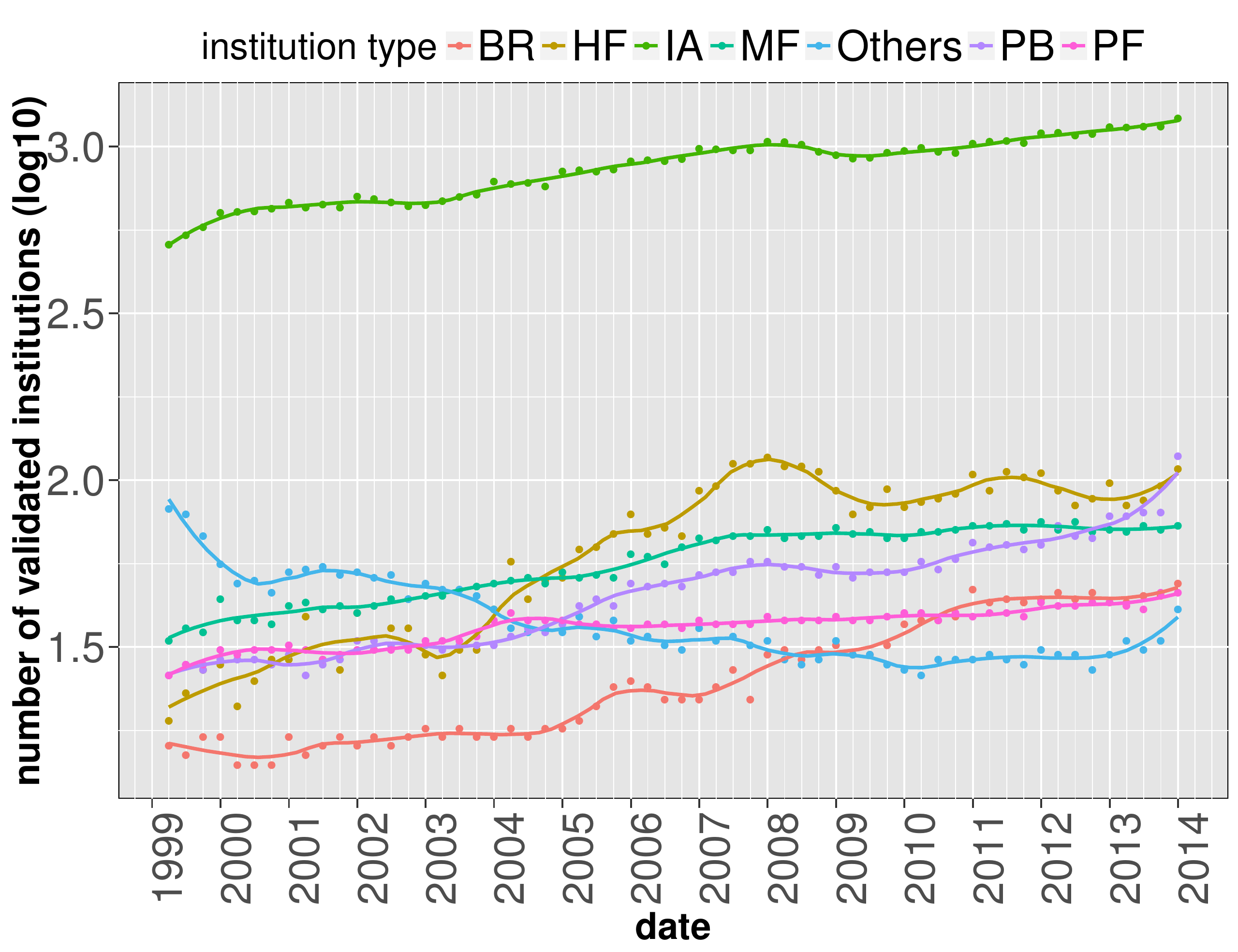}
\caption{Fraction of institutions appearing in the validated network as a function of time (top panel); 
total number of institutions in the original bipartite network (bottom left panel) and number of validated institutions (bottom right panel) for the different institution types.
Solid lines correspond to a locally weighted least squares regression (loess) of data points with 0.2 span.}
\label{fig:validated_frac}
\end{figure}

After these preliminary observations, we move to the temporal analysis of 
the structural properties of the whole validated network of institutions.
In Fig.~\ref{fig:validated_frac} we show the fraction of validated institutions (defined as the number of institutions having at least one validated link 
over the total number of institutions appearing in the ownership network) as a function of time. We also disaggregate data according to the type of institution 
and plot in this case both the number of validated institutions and the original number of institutions (we avoid to use directly their ratio for a better visualization). 
One sees that there is no particular pattern and the fraction of validated institutions is almost constant in time.  
By looking at disaggregated data, a few interesting things emerge. Investment Advisors account for the largest percentage of institutions and, more prominently, of validated institutions, 
followed by Hedge Funds and Mutual Funds. The most interesting behavior is however that of Hedge Funds in the validated networks: 
they are relatively under-represented until 2004, but after that their number displays a steep increase.

\begin{figure}[t!]
\centering
 \includegraphics[scale=0.25]{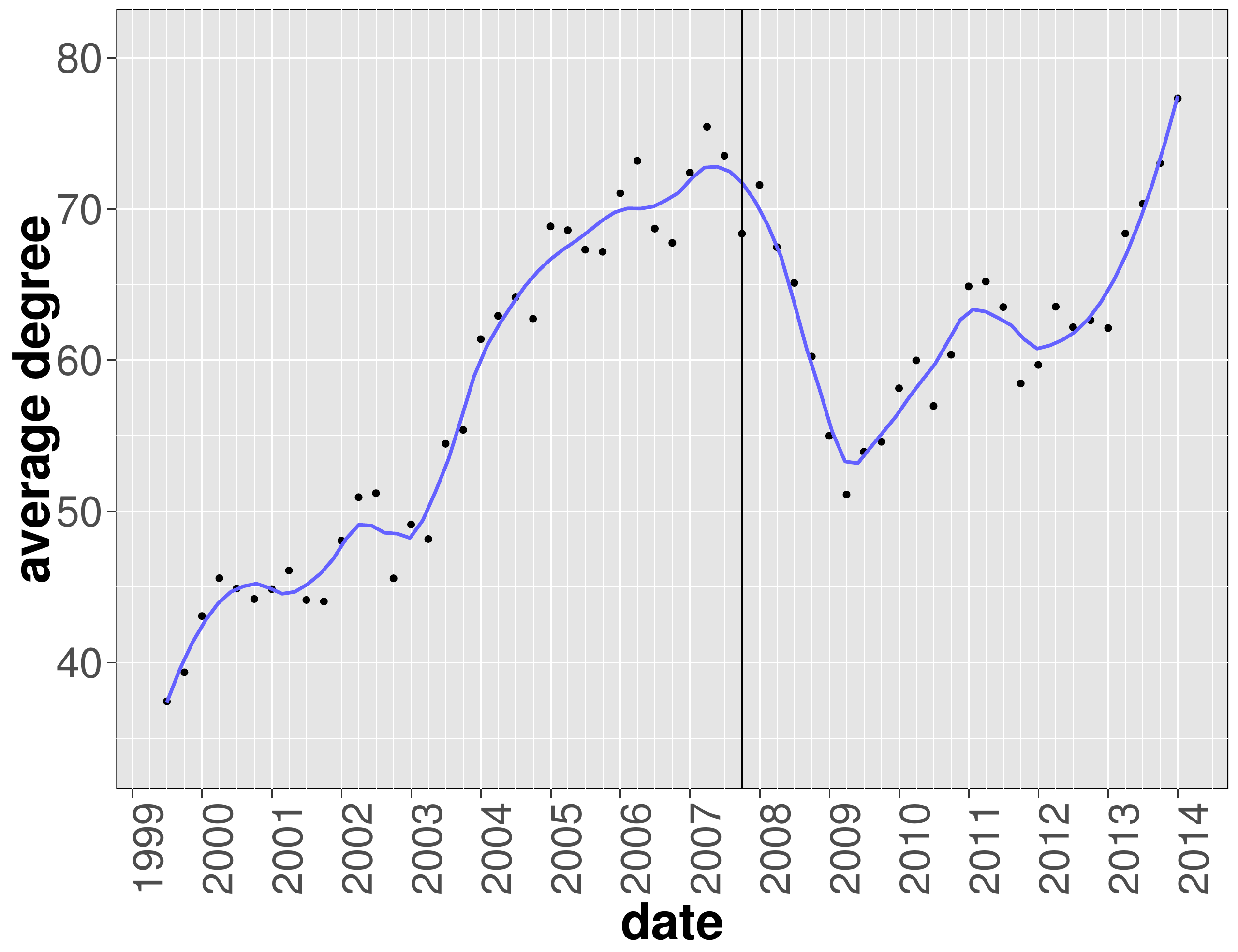}
 \includegraphics[scale=0.25]{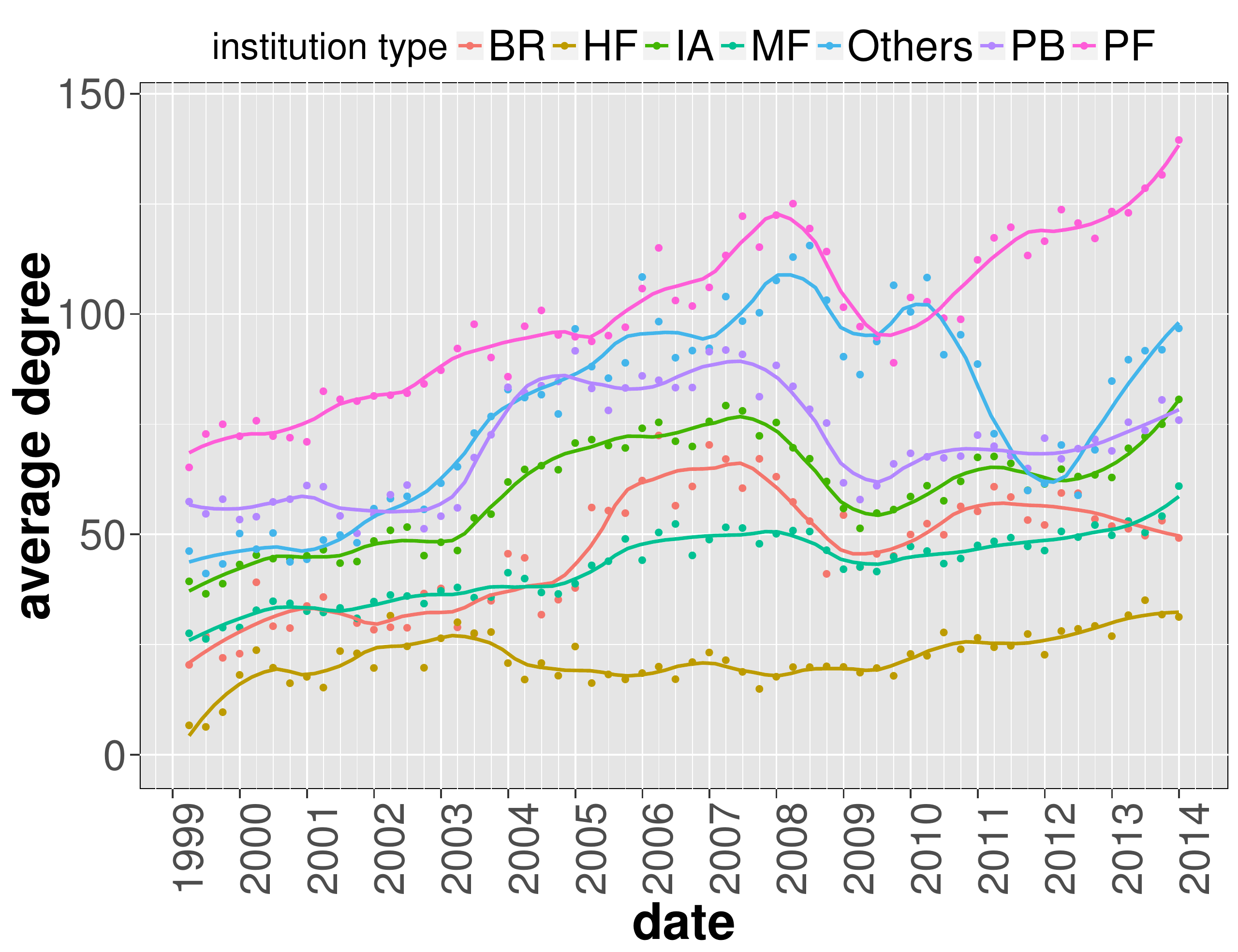}
\caption{Average degree of institutions in the validated network  as a function of time, aggregated (left panel) and separated for the different institutions type (right panel).
The vertical line correspond to the date in which we observe the maximum total market value in the dataset just before prices started to fall during the financial crisis 
(see Fig.~\ref{fig:general_info}). Remarkably, we observe a slow but steady build-up of portfolios similarity with a clear acceleration in the years 
preceding the financial crisis and from 2009 onwards. Solid lines correspond to a locally weighted least squares regression (loess) of data points with 0.2 span.}
\label{fig:validated_kavg}
\end{figure}

Fig.~\ref{fig:validated_kavg} displays the temporal evolution of the average degree in the validated network, which measures how much validated institutions are connected to each other.
One clearly sees an overall increasing trend with a strong acceleration during the years preceding the financial crisis. In particular, the average degree reaches a maximum few months before prices started to fall. 
Furthermore, our results suggest that a similar process is taking place after 2009, a fact that might question the stability of financial markets nowadays. 
The right-hand side plot of Fig.~\ref{fig:validated_kavg} reports the same quantity for each category of institutions, which also has peaks just before the 2008 crash. 
The notable exception is Hedge Funds, whose average degree is roughly constant in time. In addition, the peak for Investment Advisors, Private Banking funds and Brokers occurs roughly 1-2 quarters before the global peak.

\subsection*{Validated overlaps vs portfolio size and security capitalization}

A seemingly major shortcoming of using a binary holding matrix $\mathcal{A}(t)$ for validation purposes is that of not taking into account neither the concentration of ownership of a given security 
({\em i.e.}, which fraction of the outstanding shares a given institution is holding) nor the relative importance of different securities in a portfolio 
({\em i.e.}, which percentage of the total portfolio market value a security is representing). These are clearly important types of information, since one would expect a mild price impact 
following the liquidation of the asset by an institution if the latter owns only a small fraction of that security's outstanding shares. 
Conversely, if the asset represents a considerable fraction of the portfolio market value, a price drop will have a stronger impact on the balance sheet of the institution.
However, despite validating weighted overlaps $o^W_{ij}(t)=\sum_sW_{is}(t)W_{js}(t)$ is more relevant than binary overlaps to identifying fire sales propagation channels, 
we cannot use the original weighted matrix $\mathcal{W}(t)$ in the validation procedure, as in this case it would be impossible to build an analytical null model---which would make the validation procedure extremely involving. 
Thus, we are forced to rely on binary overlaps. However, the dataset at our disposal allows us to check a posteriori the features of the portfolio positions which contributed to the formation of validated links.

\begin{figure}[t!]
\centering
\includegraphics[scale=0.25]{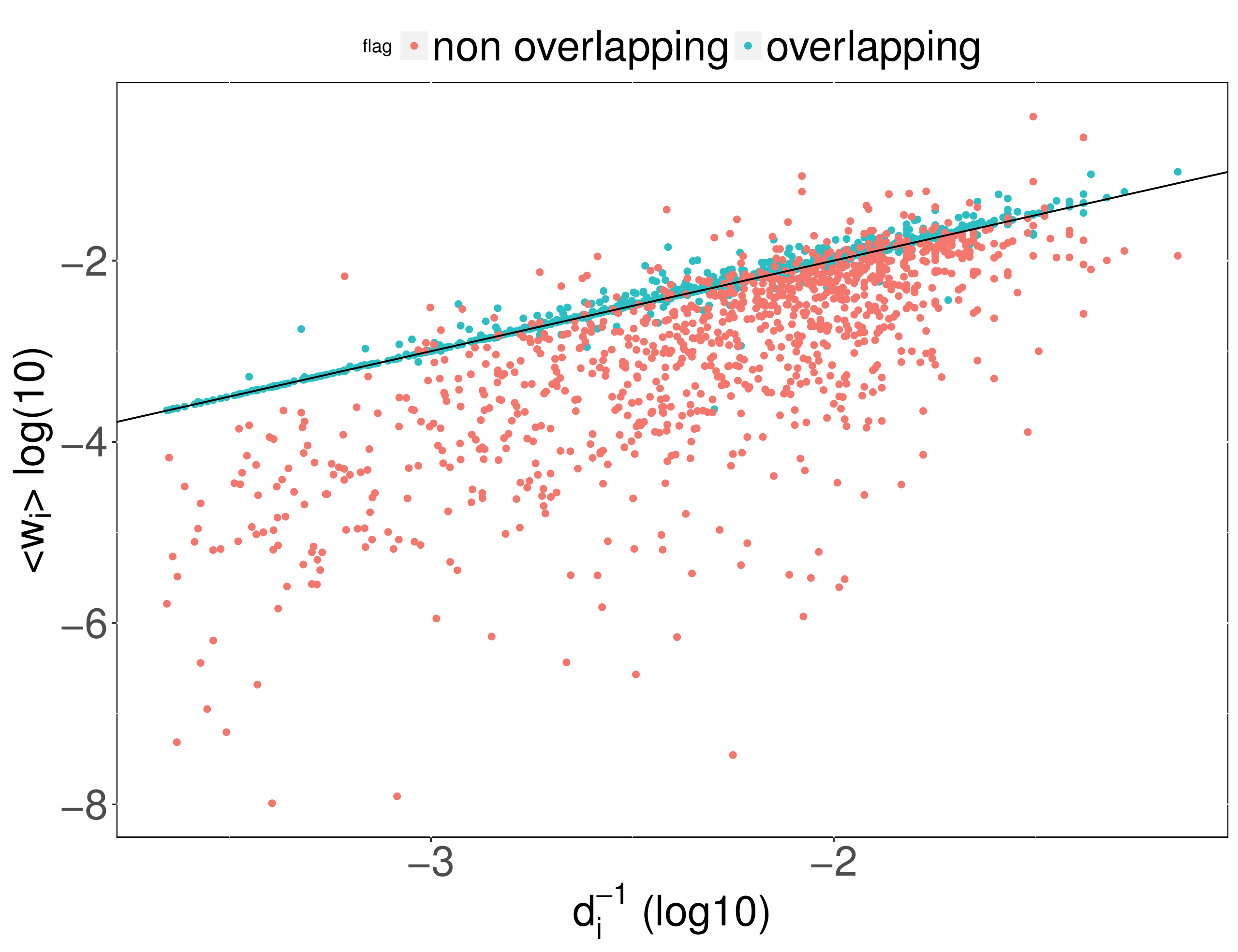}
\caption{Scatter plot of the average share of securities market value in a portfolio $\avg{w_i(t)}=\sum_sw_{is}(t)$ versus the inverse of the portfolio diversification $1/d_i(t)$ for each institution $i$. 
The average over all securities in a portfolio gives, by construction, the inverse of the institution's degree (corresponding to the straight line in the plot). 
Here we divide the average share over overlapping securities ({\em i.e.}, securities in the portfolio belonging to the overlap with a validated neighbor) 
and non-overlapping securities (the complementary set). We clearly see that overlapping positions correspond to larger shares in the portfolio.
The plot refers to 2006Q4, yet the same qualitative behavior is observed for other dates.}
\label{fig:fund_scatter}
\end{figure}

To this end, using the information about the price $p_s(t)$ and outstanding shares $\sigma_s(t)$ of different securities $s$ at time $t$, 
we compute the fraction of the total market value of portfolio $i$ represented by security $s$, 
namely $w_{is}(t)=p_s(t)W_{is}(t)/\sum_{x}p_x(t)W_{ix}(t)$, and the fraction of outstanding shares of $s$ held by institution $i$, 
namely $c_{is}(t)=W_{is}(t)/\sigma_s(t)$. We apply this procedure to each position $W_{is}(t)$ of the bipartite ownership network 
in order to characterize the features of the positions belonging to validated overlaps. 
Fig.~\ref{fig:fund_scatter} shows that, on average, overlapping securities ({\em i.e.}, securities making up the validated overlaps) represent a larger share of the validated portfolio, namely $6\%$ more than the 
average share given by the inverse of the degree. 

\begin{figure}[t!]
 \centering
\includegraphics[scale=0.19]{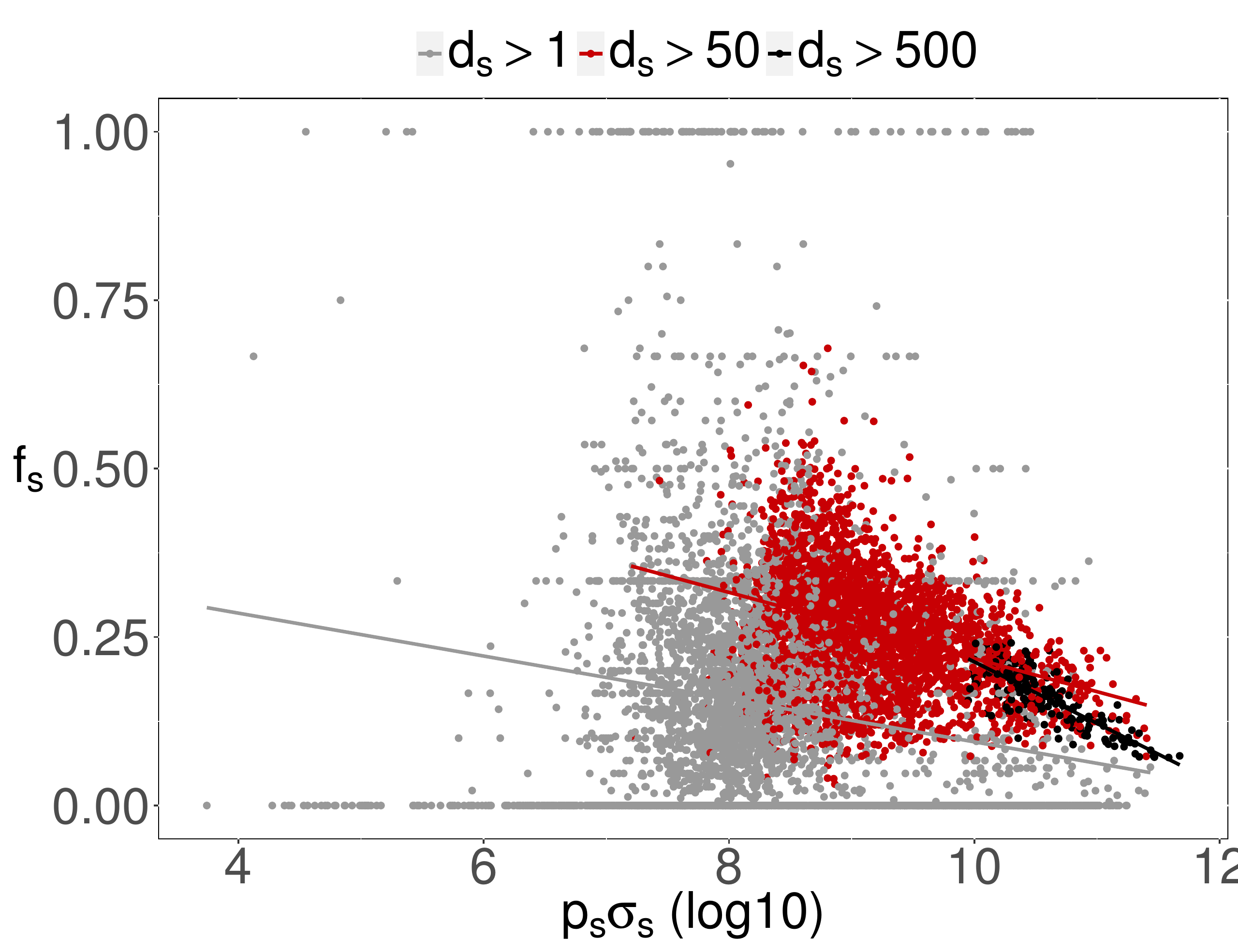}\quad
\includegraphics[scale=0.19]{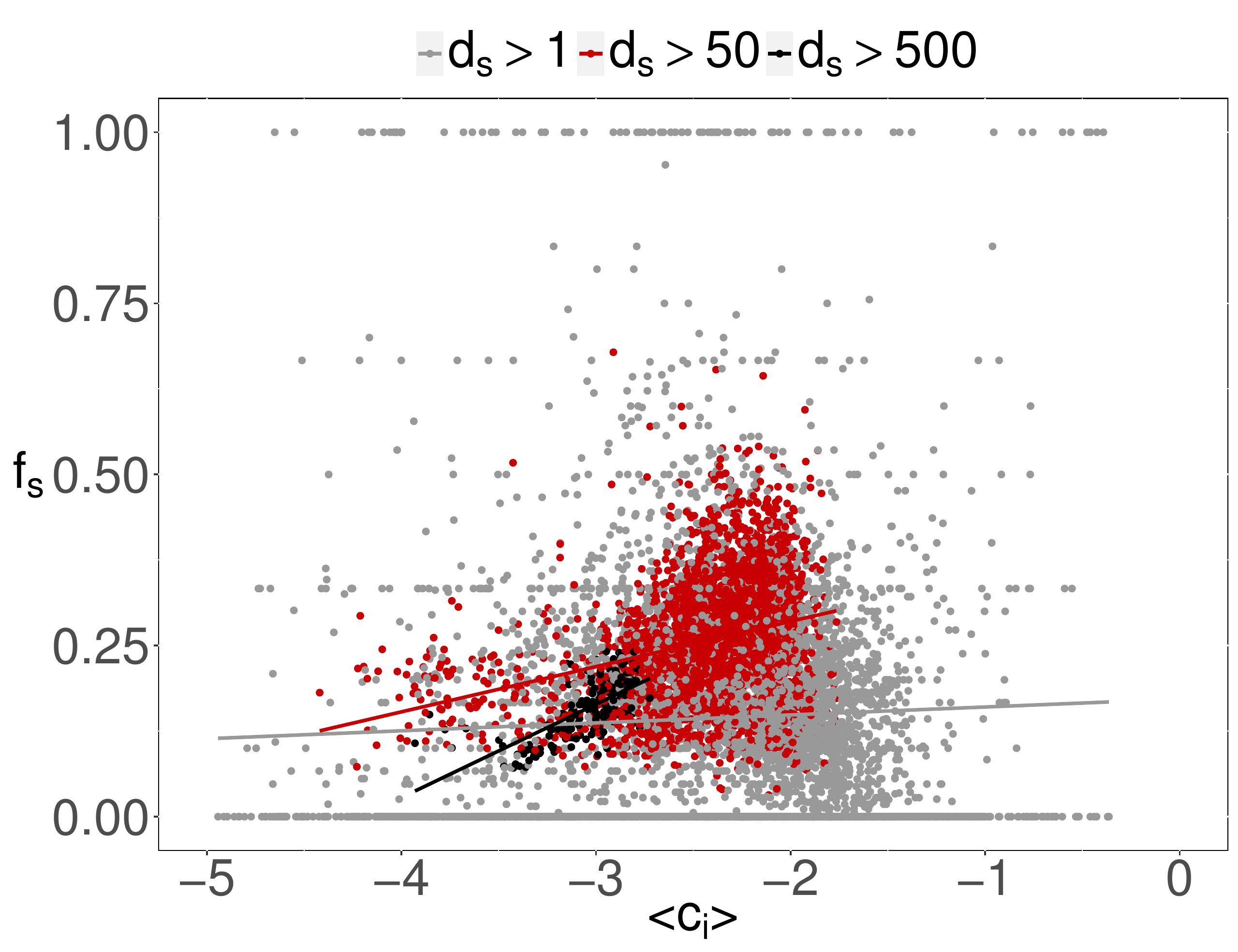}\quad
\includegraphics[scale=0.19]{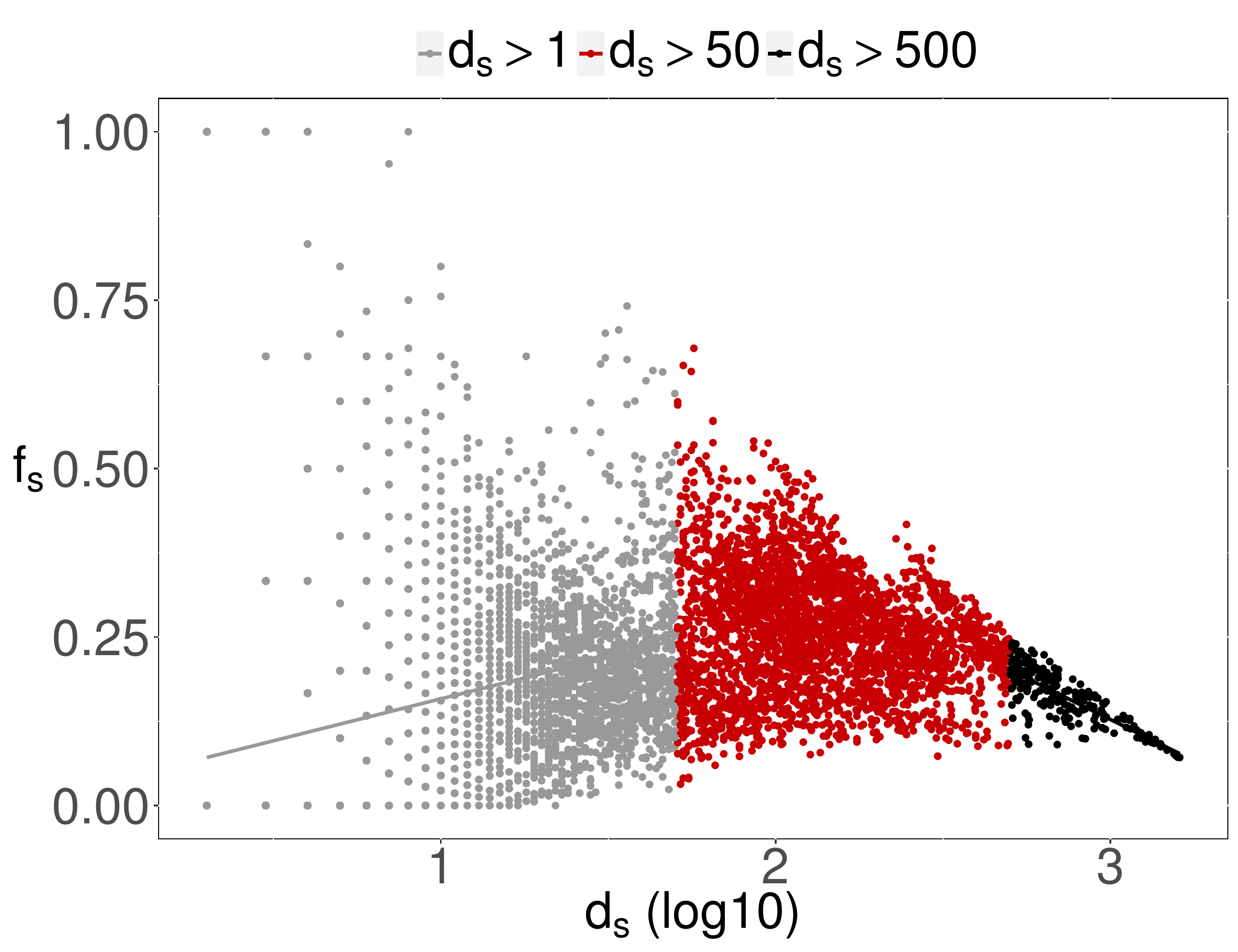}
\caption{Scatter plots of the fraction $f_s(t)$ of validated pairs of institutions owning a security $s$ versus the security capitalization (left plot), its concentration (central plot) and number of owners (right plot). 
The straight lines show log-linear regressions of the data, divided according to securities degrees. Plots refer to 2006Q4, yet the same qualitative behavior is observed for other dates.}
\label{fig:cap_scatter}
\end{figure}

In order to study the concentration of ownership of different securities we use the following procedure. Each security $s$ belongs by construction to $d_s(t)[d_s(t)-1]/2$ pairs of overlapping portfolios, 
and we can compute which fraction $f_s(t)$ of such pairs that are validated by the algorithm. We then compute for each security the total capitalization (as a proxy for the liquidity of the security) 
as well as the average ownership fraction per institution $\avg{c_i(t)}=\sum_ic_{is}(t)/d_s(t)$ as a function of $f_s(t)$. In Fig.~\ref{fig:cap_scatter} we show scatter plots of this quantities together 
with straight lines obtained from log-linear regressions. As one can see, the probability that any pair of institutions investing in the same asset are validated by the algorithm 
decreases as a function of the capitalization of the asset, increases as a function of the concentration ({\em i.e.}, with the average fraction of outstanding shares detained by an institution) 
and decreases as a function of the degree of the security. The relation is stronger for securities with higher degree, because of the larger number of available data points.    

\subsection*{Distressed institutions in the validated networks}

As a final test of the effectiveness of the validation procedure, we study the ability of the algorithm to retrieve (pairs of) institutions which are about to suffer significant losses.
The dataset at our disposal indeed covers periods of financial distress (in particular the 2008 financial crisis) and it is in such periods that one would expect some institutions to incur fire sales. 
Then, if the algorithm does filter information in a useful way, the presence of a validated link between two institutions should represent a channel for the propagation of losses. 
Note that we do not attempt here to design a test for detecting self-reinforcing fire sales. Rather, we check if the presence of a validated link ultimately contains information on the occurrence of losses. 

\begin{figure}[t!]
 \centering
\includegraphics[scale=0.3]{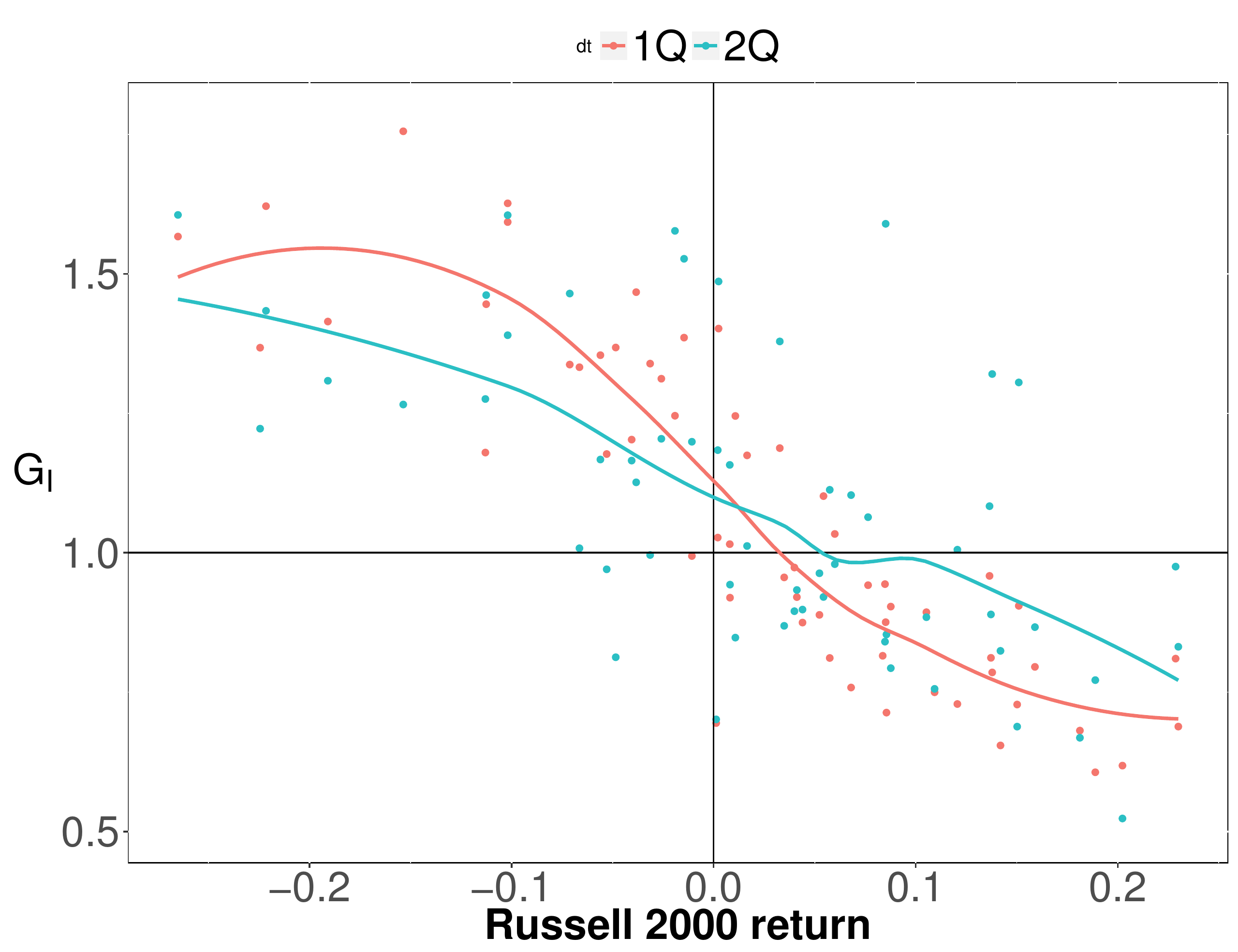}
\includegraphics[scale=0.3]{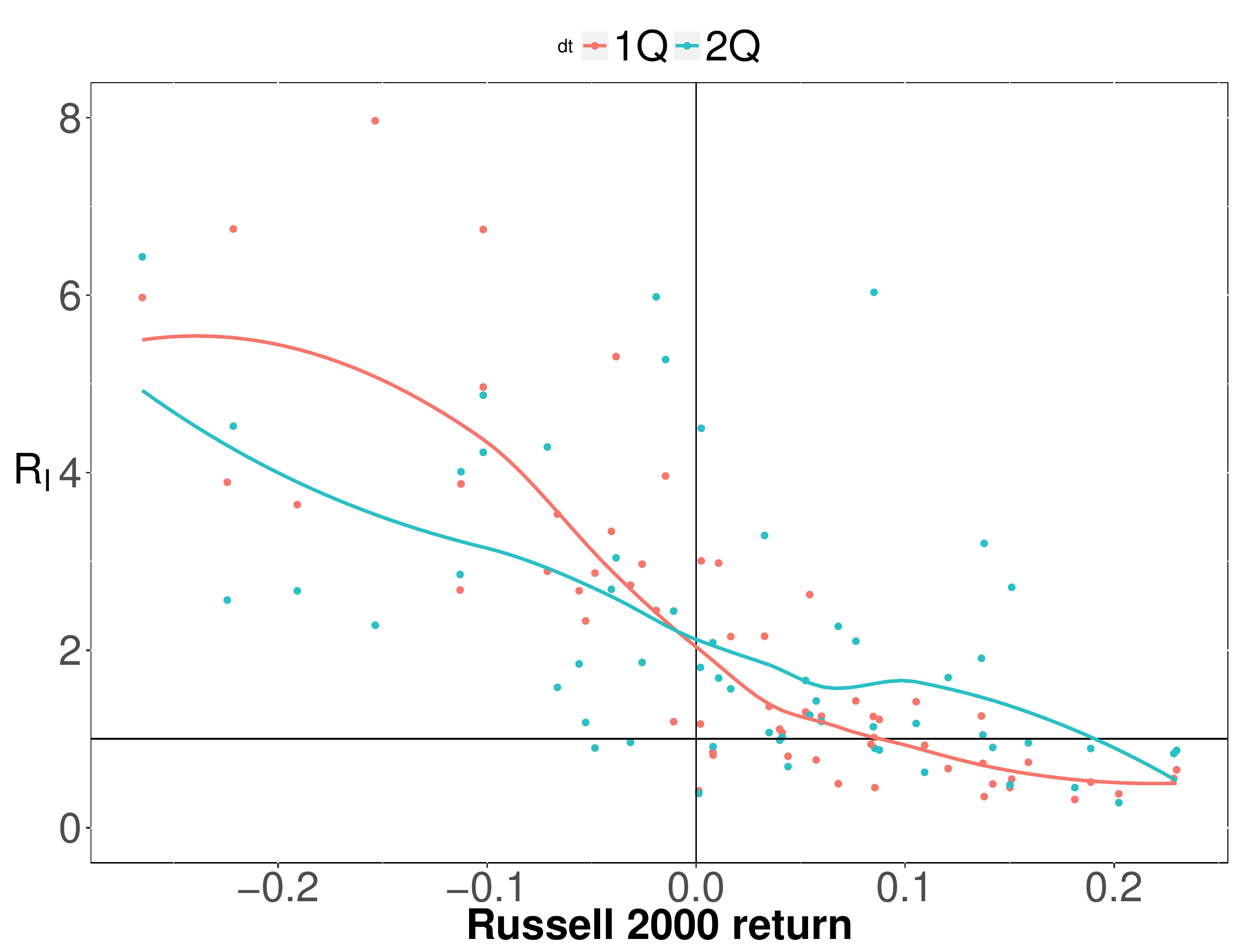}
\caption{Scatter plots of the ratios $G_I$ (left panel), {\em i.e.}, the ratio between the probability of observing a distressed institution in the validated network 
and the {\em a-priori} probability of observing a distressed institution, and $R_I$ (right panel), {\em i.e.}, the ratio between the probability of observing a linked pair of distressed institutions 
in the validated network and the probability of observing a distressed pair of institutions when all overlapping portfolios are considered, versus the return $r(t)$ between $t$ and $t+dt$ of the Russell 2000 index. 
Red points correspond to $dt$ equal to one quarter, blue points to $dt$ equal to two quarters; solid lines correspond to a locally weighted least squares regression (loess) of data points with 0.2 span. 
Panels are divided in four regions, corresponding to probabilities larger/smaller than one ({\em i.e.}, distressed institutions over/under represented in the validated networks) 
and to $r(t)$ larger/smaller than zero ({\em i.e.}, market contraction/growth).}
\label{fig:guess_scatter}
\end{figure}

To this end, we construct for each date $t$ the set $\mathcal{L}_n(t)$ of the $n$ institutions experiencing the highest drop in portfolio value between $t$ and $t+dt$ 
(which we refer to as ``distressed'' institutions). We first consider drops in absolute terms ({\em i.e.}, the total dollar amount) which we believe is of macroeconomic significance and check the relation with portfolio returns later on.
We use here $n=300$ (roughly corresponding to $10\%$ of the total number of institutions) and omit in the following the $n$ subscript. 
We then compute the fraction $l(t)$ of distressed institutions with respect to the total number of institutions $I(t)$ and compare it with the fraction of distressed institution $l_\mathcal{V}(t)$ in the validated network. 
The ratio $G_I(t)=P[i\in\mathcal{L}(t)\,|\,i\in \mathcal{V}(t)]/P[i\in \mathcal{L}(t)]=l_\mathcal{V}(t)/l(t)$ then indicates if distressed institutions are over-represented in the validated network. 
Indeed, if $G_I(t)=1$ the algorithm is not doing anything better than putting distressed institutions at random in the validated network, 
whereas, if $G_I(t)>1$ we effectively gain information by knowing that a institution belongs to $\mathcal{V}(t)$. 
Similarly, we compare the fraction of links in the validated network which connect institutions that are both distressed with the fraction of such links when all overlapping pairs of institutions 
({\em i.e.}, all pairs whose portfolios having at least one security in common) are considered. The ratio between these two quantities, namely 
$R_I(t)=P[i,j\in\mathcal{L}(t)\,|\,V_{ij}(t)=1]/P[i,j\in\mathcal{L}(t)|o_{ij}(t)>0]$, can then be used to assess the effectiveness of the algorithm to establish a link between two distressed institutions in the validated network. 
Since all the positions in our dataset are long positions, it makes sense to relate $G_I(t)$ and $R_I(t)$ to an index that encompasses many securities. 
Fig.~\ref{fig:guess_scatter} shows these quantities as a function of the market return $r(t)$ between $t$ and $t+dt$ as measured by the Russell 2000 index. 
Indeed, both ratios are correlated with the total loss, and are significantly larger than 1 when $r(t)\ll 0$ ($R_I$ in particular reaches values close to 8 in periods of major financial distress).
Notably, both ratio are close to 1 when the market loss is close to 0, and decline afterwards. This could be interpreted as the fact that, in times of market euphoria, 
overlapping portfolios turn into self-reinforcing bubbles. 

\begin{figure}[t!]
 \centering
\includegraphics[scale=0.28]{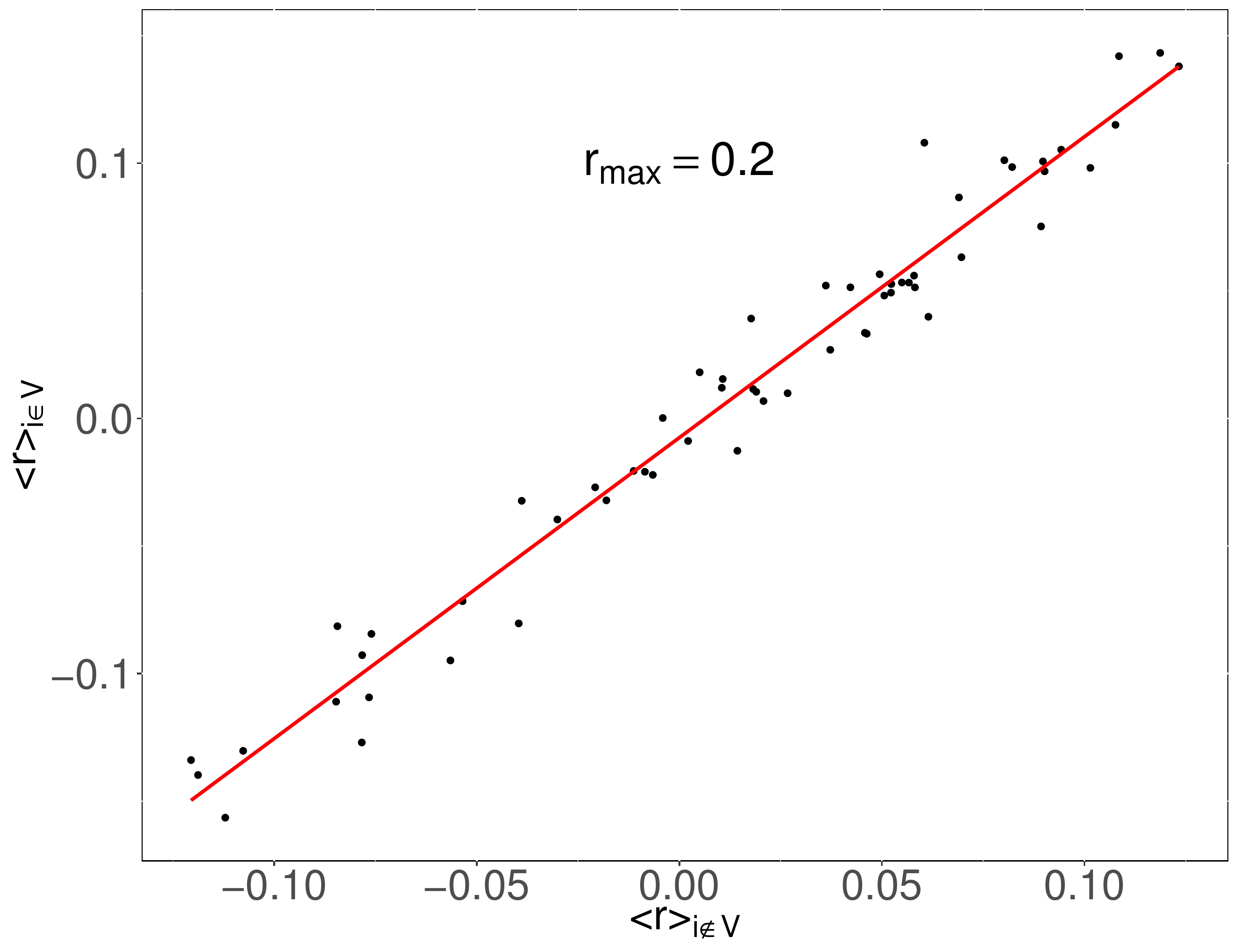}\quad
\includegraphics[scale=0.32]{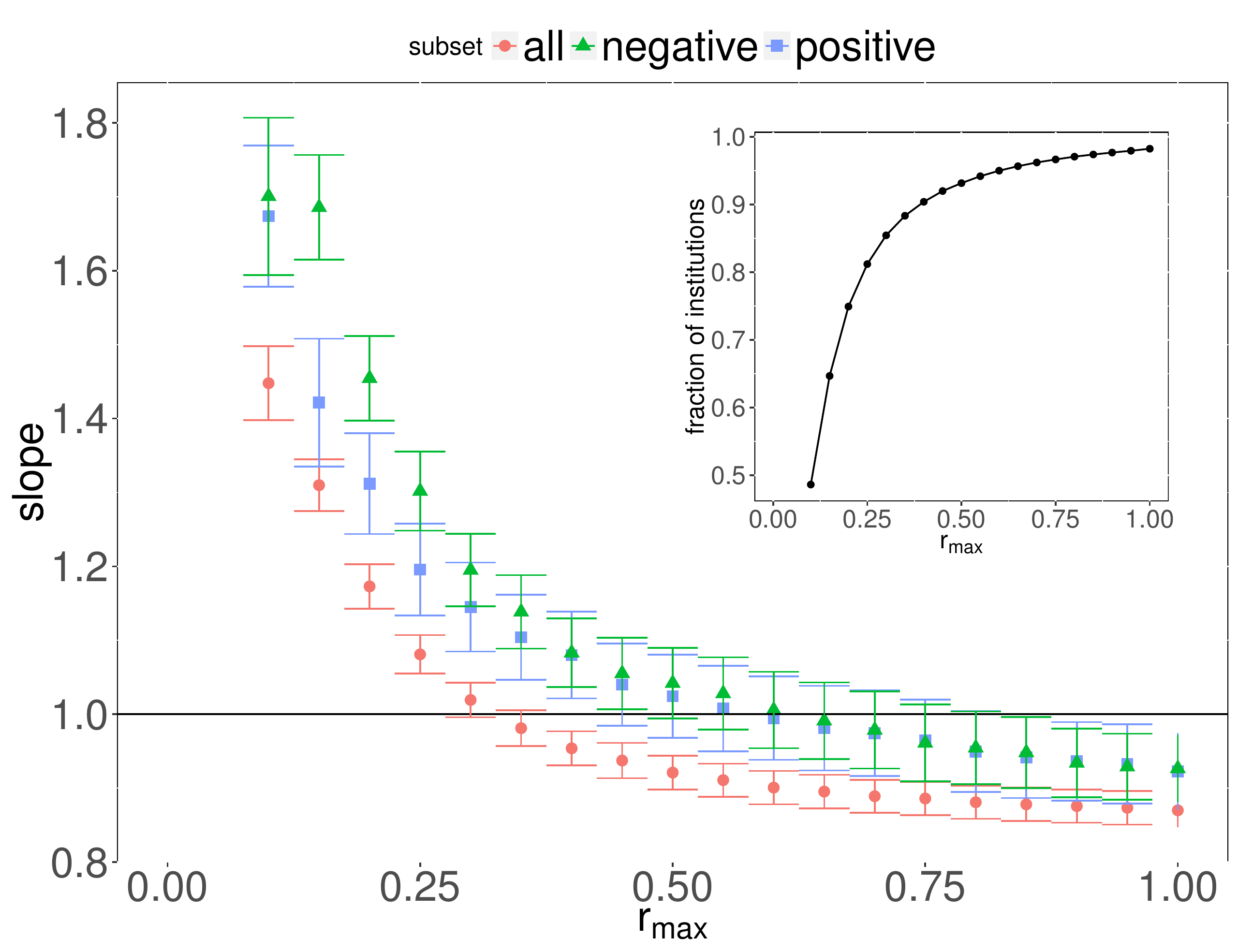}
\caption{\emph{Left panel:} average return of portfolios in the validated network vs average return in the complementary set. All returns for which $|r_i(t)|<r_{max}$ (here 0.2) are included. 
The straight line correspond to a linear regression of the data-points (one for each quarter). \emph{Right panel:} value of the slope obtained as in the left panel as a function of $r_{max}$ (see details in the text). 
When returns greater than roughly $30-40\%$ are excluded the slope is found significantly larger than $1$. This indicates that portfolios in the validated network tend to have higher returns (in absolute terms) 
than their not validated counterpart. The inset shows the overall fraction of returns satisfying $|r_i(t)|<r_{max}$ as a function of $r_{max}$.}
\label{fig:returns}
\end{figure}

When we repeat the same procedure for portfolio returns ({\em i.e.}, we use portfolio returns to label institutions as distressed) we do not obtain meaningful results. This is however due to the fact that abnormal returns 
are in general observed for small portfolios for which we have few data points. Given the statistical nature of our method we cannot hope to correctly identify such situations for which a different (probably case by case) 
methodology is clearly needed. We can however take a simpler point of view and take for each time $t$ all portfolios whose return is smaller (in absolute term) than a threshold $r_{max}$ that we use as a parameter. 
We then use this subset to compute the average return of validated portfolios $<r>_{i\in\mathcal{V}(t)}(r_{max},t)$ together with the average return of portfolios outside the the validated 
network $<r>_{i\notin\mathcal{V}(t)}(r_{max},t)$. For a given value of $r_{max}$, we then have a scatter plot of these two quantities (one point for each date $t$) 
which is well approximated by a straight line (see Fig.~\ref{fig:returns} left panel for an example). Note that with the 
significance threshold $P^*(t)$ used one has roughly half of the institutions in each set (see Fig.~\ref{fig:validated_frac} left panel).
Finally, we linearly regress  $<r>_{i\in\mathcal{V}(t)}(r_{max},t) = A<r>_{i\notin\mathcal{V}(t)}(r_{max},t)+B$, and plot the value of the slope as a function of the threshold $r_{max}$.
As one can see in the right panel of Fig.~\ref{fig:returns}, the slope is significantly larger than $1$ for values of the threshold up to roughly $30\%$ in general, and up to $50\%$ 
when we consider positive and negative returns separately. In the latter case we first split for each date institutions with positive/negative returns and 
compute return averages in the validated and complementary set. The fact the the slope become slightly smaller than $1$ for large values of $r_{max}$ 
is putatively due to abnormal returns, most likely associated with small portfolios which tend to be outside the validated network.
While this drawback is unavoidable given the statistical nature of our method, on the overall these results show that as long as abnormal returns 
are not considered, the returns of validated portfolios are on average greater (in absolute terms) than those of their not validated counterpart.

\subsection*{Buy and sell networks: the case of Hedge Funds}

Before moving to the analysis of the validated network of securities, we illustrate another interesting application of our method. 
Our dataset allows us to build, for each date $t$, the {\em buy} (or {\em sell}) bipartite network, corresponding respectively to the positions acquired (or sold) by each institution between $t-dt$ and $t$: 
$A_{is}^{\mbox{\tiny{BUY}}}(t)=1$ if $W_{is}(t)-W_{is}(t-dt)>0$ and $0$ otherwise; $A_{is}^{\mbox{\tiny{SELL}}}(t)=1$ if $W_{is}(t)-W_{is}(t-dt)<0$ and $0$ otherwise. 
Validation of these bipartite networks then highlights the institutions that have updated their portfolios in a strikingly similar way.
As a case study we consider the Hedge Funds (HF) buy/sell networks, meaning that we only consider the positions bought or sold by HF (discarding all other links), 
and apply the validation procedure to these subnetworks. The focus on this particular subset of funds is motivated by the Great Quant Meltdown of August 2007, during which quantitative HF, 
in particular those with market neutral strategies, suffered great losses for a few days, before a remarkable (although incomplete) reversal (see, {\em e.g.}, \cite{khandani2007happened}). 
In addition, we wish to investigate whether HF reacted in a synchronous way at the end of the 2000-2001 {\em dot-com} bubble.

\begin{figure}[t!]
 \centering
\includegraphics[scale=0.32]{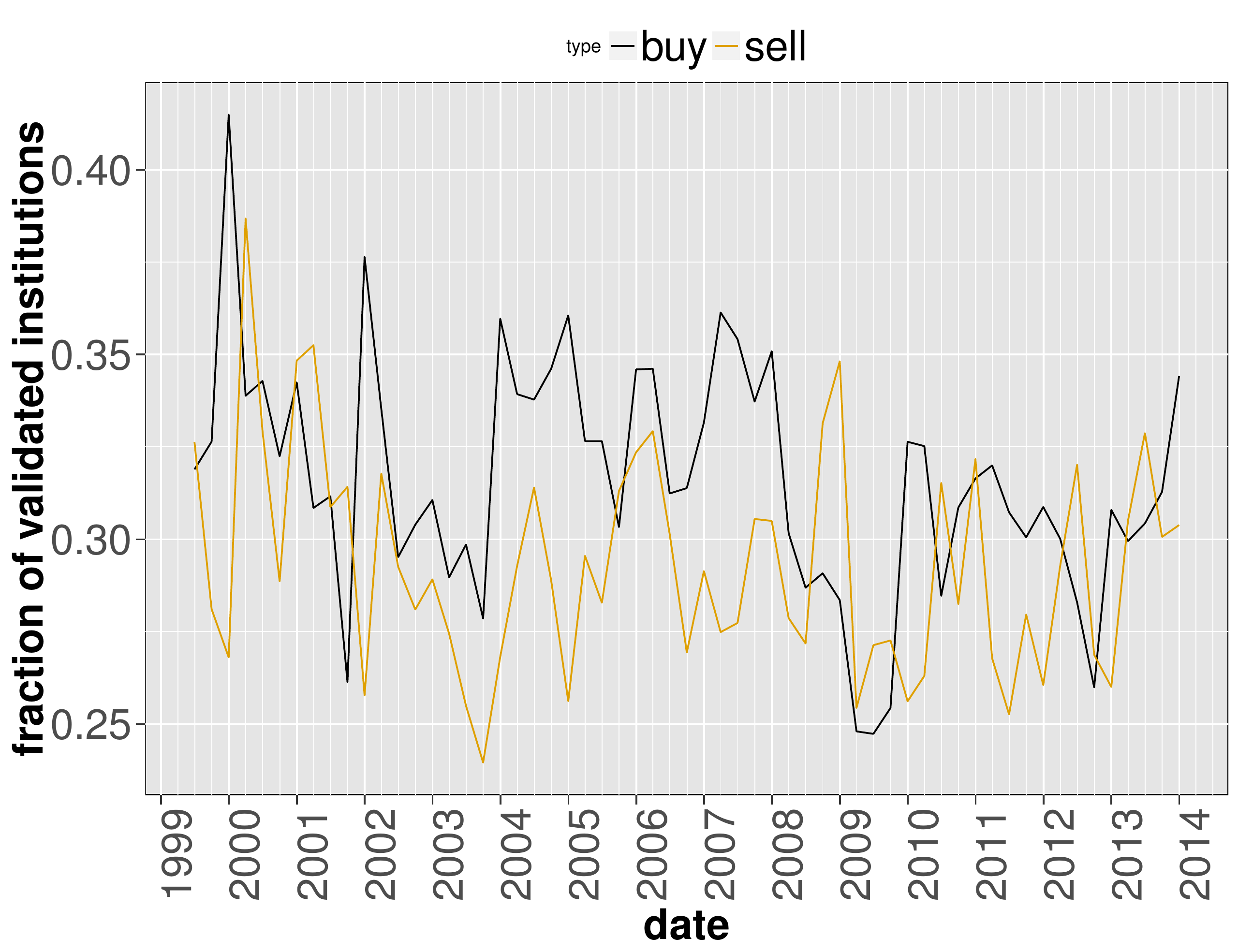}\quad
\includegraphics[scale=0.32]{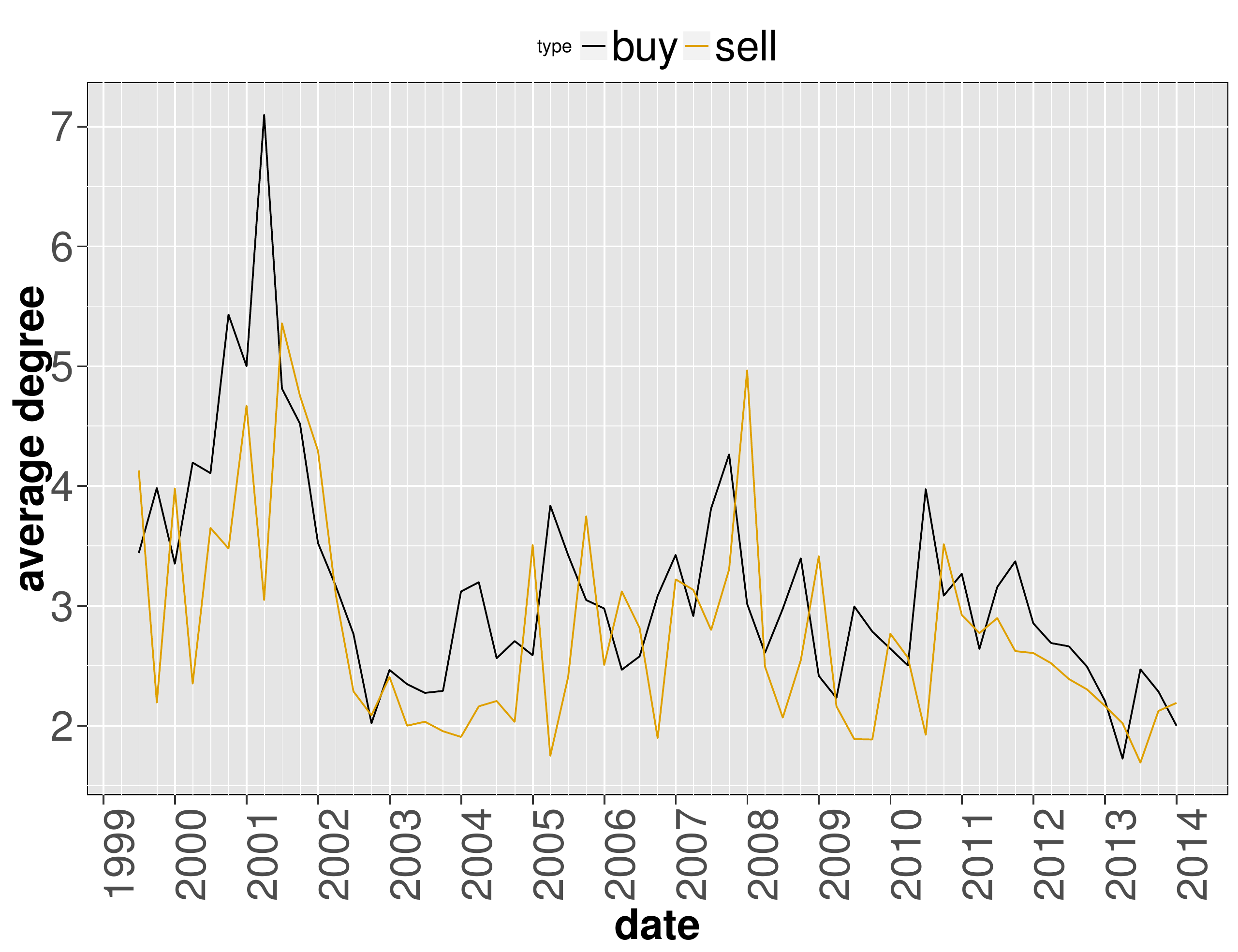}
\caption{Fraction of validated institutions (left) and average degree in the validated network (right) for the buy/sell subnetworks of Hedge Funds. 
Here the original bipartite network is made up only of Hedge Funds and the positions they acquire/sell between $t-dt$ and $t$.}
\label{fig:hedge}
\end{figure}

As for Fig.~\ref{fig:validated_fancy}, Fig.~\ref{fig:hedge} shows that the fraction of HF validated in the buy/sell network is roughly constant in time, with however some more interesting local fluctuations 
(especially in the years around $2008$). For what concerns the average number of neighbors in the validated network, one sees that the fluctuations of the sell networks lag by 3 months 
those of the buy networks: indeed, the cross-correlation is maximal at such a lag, and is quite high (0.8). This is possibly due to the fact that the typical position holding time of HF 
is smaller than 3 months: what has been bought will have been sold 3 months later. Notably, the right panel of Fig.~\ref{fig:hedge} points to the fact that buy networks are more dense on average than sell networks. 
This is also reflected in the autocorrelation of the average number of neighbors, which decrease faster for sell networks. 
Since our dataset only contains long positions, we can only conclude that HF are more synchronized when they open long positions, and liquidate them in a less synchronized way. 

Using as a first approximation the average number of validated neighbors per fund in order to assess the synchronicity of the HF actions, 
we clearly observe significant increasingly synchronized buying patterns after the top of the dot-com bubble. 
There may be two reasons for buying at this date: either the strategies of the HF were not aware of the bubble burst and were still using trend-following, 
or they took advantage of the burst to buy stocks at a discount. Noteworthy, synchronized selling lags on buying, and was overall less intense. 
Concerning the period of the global financial crisis, we observe one buy peak at 2007Q3, and one sell peak at 2007Q4. The first peak may indeed be related to the Big Quant Meltdown of August 2007. 
However, the so-called long-short market neutral funds that were forced to liquidate their positions should appear in the sell network, not the buy one. 
This would have been observed if that crisis had happened at the end of a trimester. Unfortunately, there is an almost two-month delay between the meltdown and the reporting, which probably hides the event. 
At all rates, the meltdown acted as a synchronization event, as the buy network density is clearly an outlier at the end of September 2007: 
HF have therefore acquired significantly similar long positions in their portfolios during the same quarter, and then, expectedly, liquidated them by the end of the next trimester.

\subsection*{Temporal evolution of the validated network of securities}

\begin{figure}[t!]
\centering
 \includegraphics[scale=0.3]{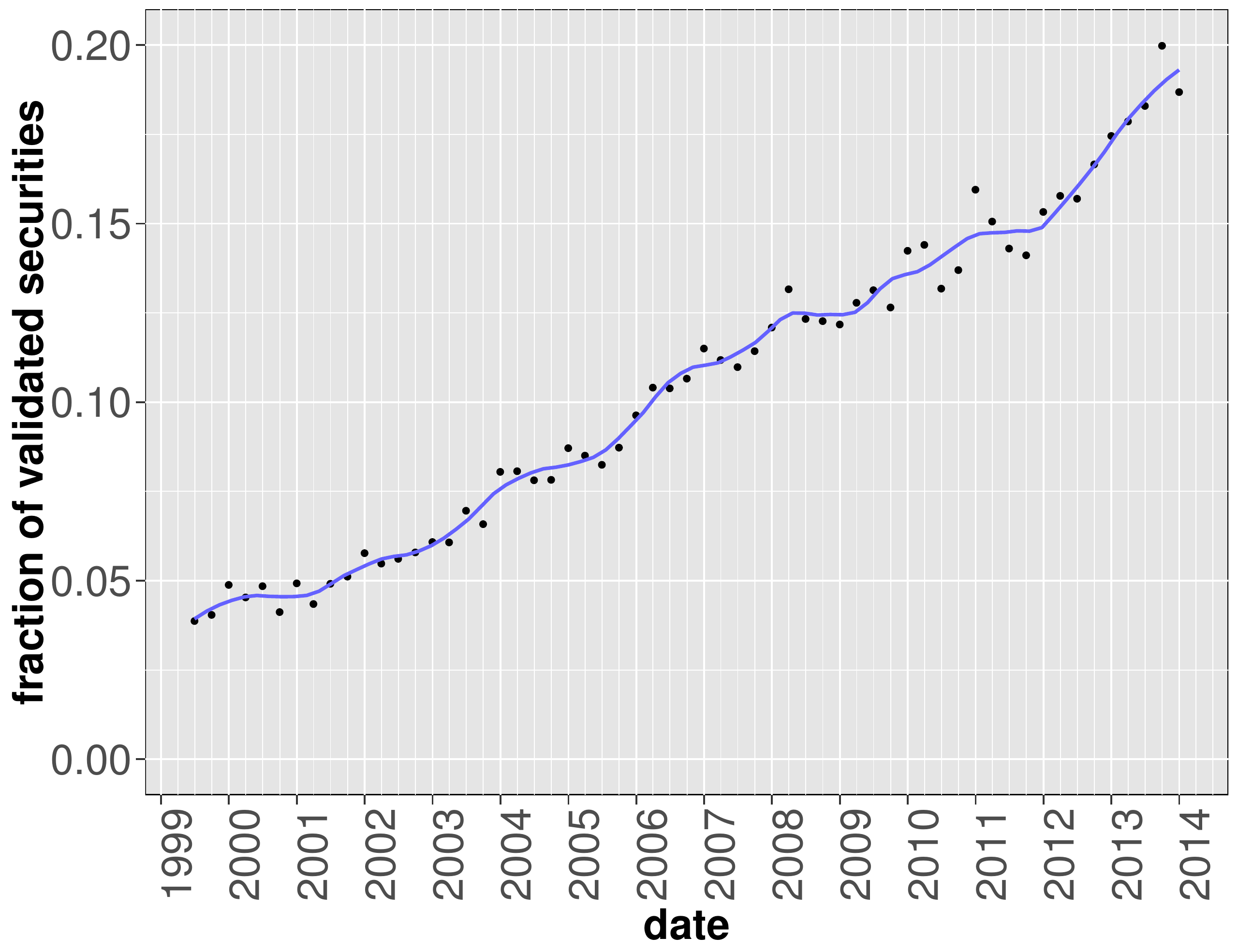}\quad
 \includegraphics[scale=0.3]{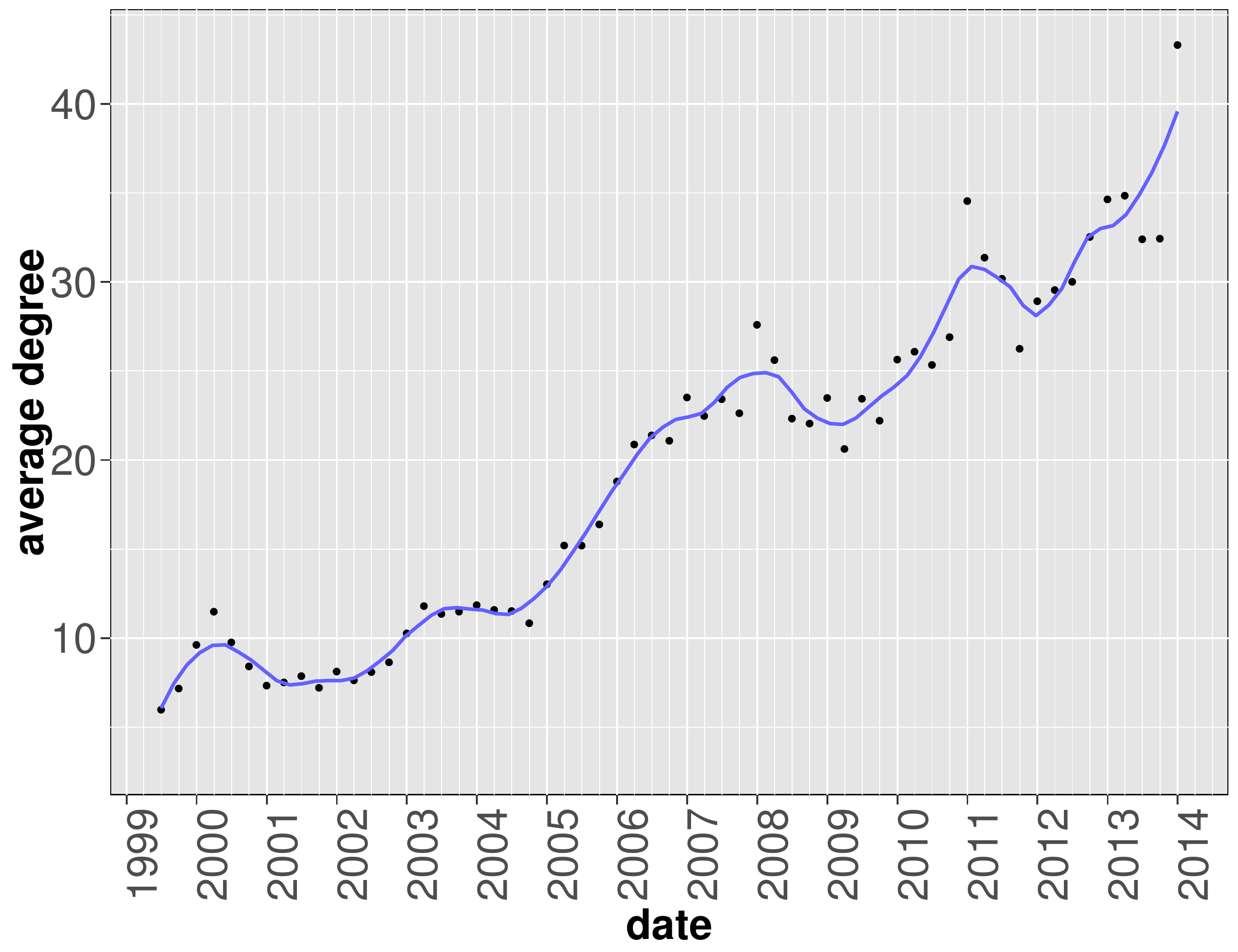}
\caption{Fraction of securities appearing in the validated network (left panel) and their average degree in the validated network (right panel) as a function of time. 
Differently from the validated network of institutions, here the number of validated securities grows steadily in time. 
Yet, the number of validated links grows at a faster peace, as demonstrated by the increasing average degree. Solid lines correspond to a locally weighted least squares regression (loess) of data points with 0.2 span.}
\label{fig:validated_securities}
\end{figure}

In this section we finally use our method to detect statistically significant common ownerships of securities, in order to identify contagion channels between securities themselves. 
Thus, we apply the validation procedure to the security ownership overlap $\tilde{o}_{sq}(t)=\sum_iA_{is}(t)A_{iq}(t)$ (instead of the institution portfolios overlap 
$o_{ij}(t)=\sum_sA_{is}(t)A_{js}(t)$). The presence of a validated link between two securities then reflects the fact that they share a significantly similar set of owners, 
which again translates into a potential contagion channel through fire sales. 
Fig.~\ref{fig:validated_securities} shows the temporal evolution of aggregate features of the validated network projection on securities. Contrarily to the case of the 
institutional projection (Fig.~\ref{fig:validated_frac} and~\ref{fig:validated_kavg}), here we observe a stable growth of validated securities: 
there are more and more stocks that can be involved in a potential fire sale (or closing down of similar institutions). 
Moreover, as testified by the growth of the average degree of validated securities, the validated network becomes denser, signaling the proliferation of contagion channels for fire sales. 
Note the presence of local maxima that correspond to all major financial crises covered by the database: the {\em dot-com} bubble of 2001, the global financial crisis of 2007-2008 
and the European sovereign debt crisis of 2010-2011. As for the case of institutions, the similarity pattern of securities ownerships is maximal at the end of the considered time span.

The fact that the average degree of the validated network of securities keeps growing boils down to the fact that institutions choose securities, not the opposite. 
While the number of institutions in our dataset has increased over the years, the number of securities has been roughly constant. 
If a new institution selects at random which assets to invest in, then the average degree of the securities network would stay constant. 
This is not the case, if only because of liquidity constraints. Therefore, on average, the portfolio of a new institution is correlated with the ones of pre-existing institutions.

\begin{figure}[t!]
\centering
 \includegraphics[scale=0.5]{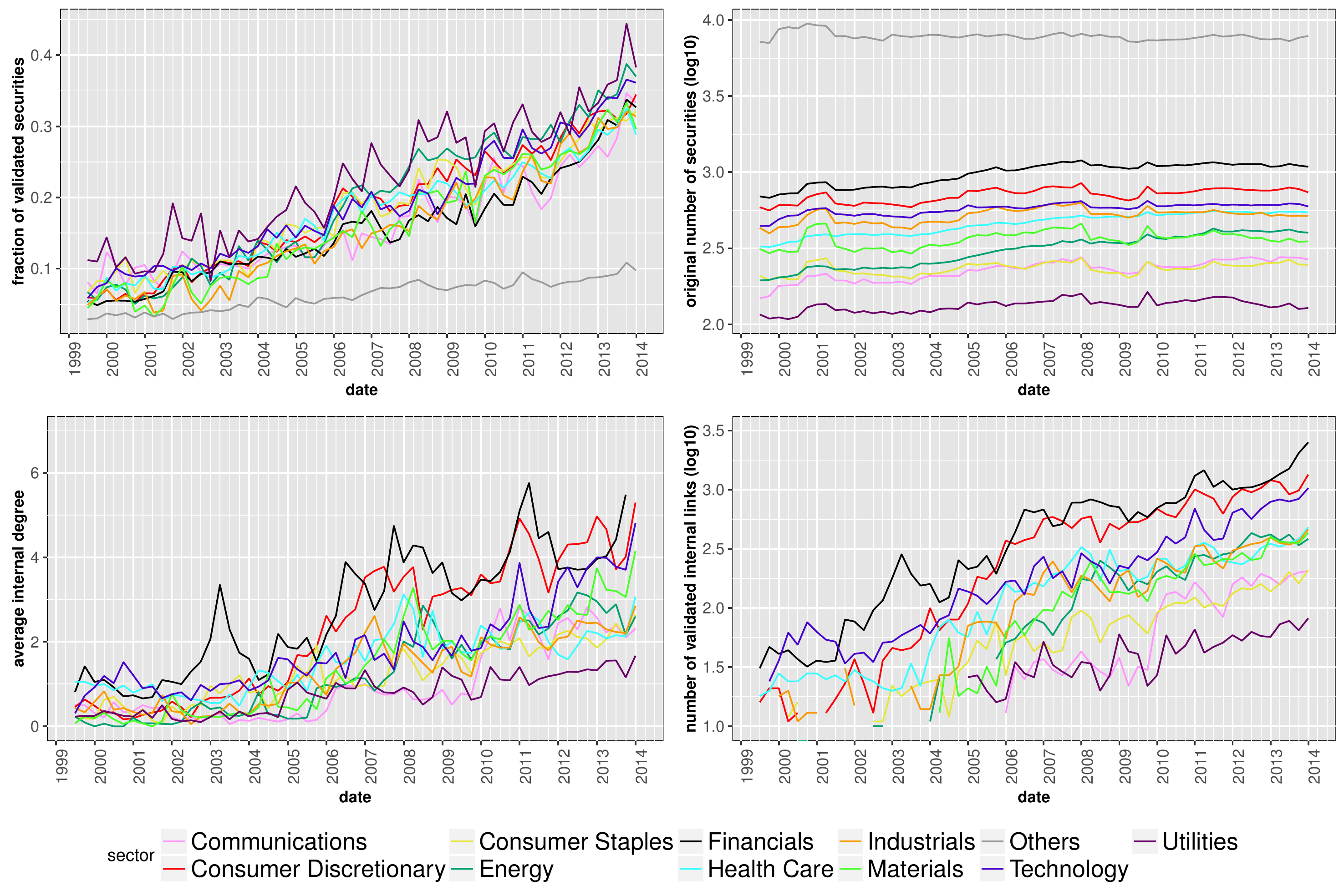}
\caption{Statistics of the validated networks of securities, disaggregated by BICS category: 
fraction of validated securities (upper left panel), total number of securities in the original bipartite network (upper right panel), 
average internal degree (lower left panel) and internal links (lower right panel) in the validated network. 
The latter two quantities are obtained by considering only validated links connecting securities of the same category. 
Security categories that are more internally connected are Financials (which includes the following level 2 sectors: 
Banking, Commercial Finance, Consumer Finance, Financial Services, Life Insurance, Property \& Casualty, Real Estate) 
and Consumer Discretionary (which includes: Airlines, Apparal \& Textile Products, Automotive, Casinos \& Gaming, Consumer Services, 
Distributors, Educational Services, Entertainment Resources, Home \& Office Products, Home Builders, Home Improvements, Leisure Products, Restaurants, Travel \& Lodging).}
\label{fig:validated_securities_sec}
\end{figure}

In order to detect if the observed patterns concern peculiar classes of securities, we perform an analysis of the validated network distinguishing securities according to the 
Bloomberg Industry Classification Systems (BICS) \cite{BICS2013}---which rests on their primary business, as measured first by the source of revenue and second by operating income, assets and market perception. 
Each security thus belongs to one of the following sectors: Communications, Consumer Discretionary, Consumer Staples, Energy, Financials, Health Care, Industrials, Materials, Technology, Utilities (or other).
In particular, we try to detect whether securities of the same category tend to be connected together in the validated network. 
To this end, we denote as {\em internal} a validated link connecting two securities with the same BICS label, and we compute the internal degree as the degree of a security restricted to internal links. 
As Fig. \ref{fig:validated_securities_sec} shows, the categories of securities that are more internally connected are (notably) Financials and, to a lesser extent, Consumer Discretionary. 
This does not mean that portfolio overlaps concentrate on these categories, but rather that relatively more contagion channels exist within securities belonging to them.

\section*{Conclusion}\label{sec:con}

In this work, we have proposed an exact method to infer statistically robust links between the portfolios of financial institutions based on similar patterns of investment. 
The method solves the problem of evaluating the probability that the overlap of two portfolios of very different size and diversification 
is due to random allocation of assets of very different market capitalization and number of owners. The use of an appropriate null hypothesis 
provided by the bipartite configuration model~\cite{Saracco2015SR} considerably improves the statistical significance of the detected features of the validated networks. 
Note that the method is general, and can be applied to any bipartite network representing a set of entities sharing common properties 
({\em e.g.}, membership, physical attributes, cultural and taste affinities, biological functions, to name a few) and where the presence of (unlikely) similar sets of neighbors is of interest 

The present study then points to the conclusion that, just before financial crises or bubble bursts, the similarity of institutions holdings increases markedly. 
Perhaps worryingly for equity markets, the proposed proxy of fire sale risk, having reached a peak in 2008 and subsequently much decreased, has been increasing again 
from 2009 to the end of our dataset (2013) up to levels not seen since 2007. Despite our method relies on binary ownership information, we also found that on average 
overlapping securities correspond to larger shares of validated portfolios, potentially exacerbating fire sales losses. 
In addition, the proposed validation method can effectively retrieve the institutions which are about to suffer significant losses in times of market turmoil 
(when validated links are the channels for which liquidation losses propagate), as well as those with the highest growth in times of market euphoria 
(when overlapping portfolios turn into self-reinforcing bubbles). Finally we show that the number of securities that can be involved in a potential fire sale 
is steadily growing in time, with an even stronger proliferation of contagion channels.

In this work we have only investigated patterns of portfolio overlap, not the probability that they lead to fire sales. This is a more complicated problem for which other datasets and econometric techniques are needed. 
However, even if we cannot draw any strong implication from our findings, all the analysis we performed confirm the coherence of our method and suggest that overlapping portfolios do play a role in financial turmoils.
Furthermore, the relationship between holdings and future portfolio changes must be better characterized. Indeed, even if two institutions with different strategies converge to a similar portfolio, 
this does not imply that they will update the latter in the same way and at the same time. 
However, it is likely that part of the institutions follow ({\em in fine}) equivalent strategies, which implies portfolio overlap and subsequent increased risk of fire sales, 
which triggers further leverage adjustment, as pointed by~\cite{Caccioli2014JBF,Cont2014MF}.
Finally, it will be useful to repeat our analysis on larger datasets so as to encompass other bubbles and crises, and to examine difference in investment patterns across various markets.

\section*{Methods}\label{sec:meths}

\subsection*{Dataset}

We extracted  13-F SEC filings (\href{url}{https://www.sec.gov/}) from the Factset Ownership database from 1999Q1 to 2013Q4, 
covering institutions valued more than 100 million dollars in qualifying assets which must report their long positions to the SEC at the end of each trimester. 
As the 13-F dataset contains only positions greater than 10000 shares or \$ 200000, very small positions are already filtered out.
The dataset is composed of a set $I(t)$ of approximately 1500$\div$3500 institutions, holding positions from a set $S(t)$ of securities, whose size fluctuates around 12500 (see Fig.~\ref{fig:general_info}). 
Note that the portfolios of sub-funds are merged into a single report. In addition to the raw ownership data, our dataset is complemented by meta-data about both institutions and securities. 

\subsection*{Significance level under multiple tests}

In order to choose an appropriate threshold (the significance level) $P^*(t)$ to 
be used in the validation procedure, we have to account for the multiple 
hypothesis tested 
(corresponding to the number $n_{pairs}(t)$ of possible pairs of institutions having a nonzero overlap). 
Here we use the rather strict Bonferroni correction~\cite{Muller1981}, meaning that we set the threshold to $P^*(t)= \epsilon/n_{pairs}(t)$. 
Note that the choice of the significance level still leaves some arbitrariness. While results presented in the paper are obtained with $\epsilon = 10^{-3}$, 
we have tested our method with various values of $\epsilon$, and employed also the less-strict false discovery rate (FDR) criterion~\cite{Benjamini1995JRSB}, without finding major qualitative differences. 
In fact, while the final size of the validated network clearly depends on the threshold, the relative temporal changes of the network statistics are much less affected by the particular value used. 

\subsection*{Resolution problems for the hypergeometric distribution approach}

As stated in the Introduction, the approach proposed in~\cite{Tumminello2011PLoS} to divide the original bipartite network into homogeneous subnetworks of securities 
has some intrinsic limitations, especially when securities are characterized by a strongly heterogeneous number of investors (as it generally happens in stock market data). 
In this circumstance, in fact, the splitting procedure often translates into almost empty subsets---especially for securities held by a large number of 
investors. 
In these subsets, overlaps can assume only a few values, bounded by the limited number of securities considered, resulting in a handful, spaced-out possible outcomes for the p-values. 
The problem then arises with the use of a global threshold corrected for multiple hypothesis testing. In fact, since institutions are compared on the many subnetworks of securities with the same degrees, 
$n_{pairs}(t)$ scales as $I^2(t)d_s^{max}(t)\equiv I^2(t)\max_s d_s(t)$: the validation threshold becomes extremely small for large and heterogeneous systems and vanishes in the infinite size limit. 
These issues lead to a serious problem of resolution, since $P^*(t)$ is too small to validate even the smallest non-zero p-value in most of the subnetworks. 
As a result, the validated network becomes almost empty by construction. Overall, while the method proposed in~\cite{Tumminello2011PLoS} works well for small networks 
with little degree heterogeneity, the same approach is not feasible in the case of large scale and highly heterogeneous networks. 

\subsection*{p-values from the Bipartite Configuration Model}

Determining the probability distributions used Eq.~(\ref{eq.pval}) requires to solve a technical problem caused by the heterogeneity of both institutions and securities. 
For example, it is hard {\em a priori} to compare a portfolio with very few assets and one with very many assets. 
However, the bipartite configuration model (BiCM)~\cite{Saracco2015SR} provides a null network model suitable for these kind of situations. 
We remand the reader to \cite{PhysRevE.70.066117,Squartini2011NJP,Saracco2015SR} for more details on the method. 
In the following we will omit the explicit time dependence of the quantities considered, since the same procedure is repeated for each date.

In a nutshell, the BiCM prescribes to build the null model simply as the ensemble $\Omega$ of bipartite networks that are maximally random, 
under the constraints that their degree sequences of institutions and security is, on average, equal to the one of the original network. 
This is achieved through maximization of the Shannon entropy of the network subject to these constraints, 
that are imposed through a set of Lagrange multipliers $\{\theta_i\}_{i=1}^I$ and $\{\theta_s\}_{s=1}^S$ (one for each node of the network). 
Solving the BiCM means exactly to find these multipliers, that quantify the abilities of nodes to create links with other nodes. 
Thus, importantly, nodes with the same degree have by construction identical values of their Lagrange multipliers. 
Once these multipliers are found, the BiCM prescribes that the expectation values within the ensemble of the network matrix element $\langle A_{is}\rangle_\Omega$, 
{\em i.e.}, the ensemble probability $Q_{is}$ of connection between nodes $i$ and $s$, is given by:
\begin{equation}
 \langle A_{is}\rangle_\Omega\equiv Q_{is}=\frac{\theta_i\theta_s}{1+\theta_i\theta_s},
\end{equation}
and the probability of occurrence $\mathcal{Q(A)}$ of a network $\mathcal{A}$ in $\Omega$ is obtained as the product of these linking probabilities $Q_{is}$ 
over all the possible $I\times S$ pairs of nodes. In other words, links are treated as independent random variables, by defining a probability measure where links correlations are discarded. 
The key feature of the BiCM model is that the probabilities $\{Q_{is}\}$ can be used to directly sample the ensemble of bipartite graphs and to compute the quantities of interest analytically. 
We can thus use the matrix $\mathcal{Q}$ to compute the expectation values of portfolios overlap between two institutions $i$ and $j$ as:
\begin{equation}
 \langle o_{ij}\rangle_\Omega = \sum_{s\in\mathcal{S}}Q_{is}Q_{js},
\end{equation}
or to compute the probability distribution $\pi(\cdot|d_i,d_j)$ of the expected overlap under the null hypothesis of random connections in the bipartite network---which, 
according to the BiCM prescription, only depends on the degrees of institutions $i$ and $j$. 
Indeed, $\pi(\cdot|d_i,d_j)$ is actually the distribution of the sum of $S$ independent Bernoulli trials, each with probability $Q_{is}Q_{js}$. 
This distribution can be computed analytically using a Normal approximation of the Poisson-Binomial distribution~\cite{Hong2013CSDA}. 
This approach has been developed by~\cite{Saracco2016arXiv} in parallel with our research. Here we discuss instead an exact and optimized numerical technique to compute $\pi(\cdot|d_i,d_j)$. 
Indeed, the computational complexity of the numerics can be substantially reduced by recalling, again, that $Q_{is}\equiv Q_{is'}$ if $d_s\equiv d_{s'}$ 
$\forall i$: 
connection probabilities only depends on nodes degree values. This is an important observation, which translates into the following statement: 
the expected overlap between any two institutions $i$ and $j$ restricted to the set of securities with a given degree 
follows a binomial distribution with probability $Q_{is}Q_{js}$ (where $s$ is one of these securities) and number of trials equal to the cardinality of such set. 
More formally, if $\{\tilde{d}_h\}_{h=1}^{d_s^{max}}$ denotes the set of {\em different} degrees within securities, 
$\tilde{n}_h$ is the number of securities having degree $\tilde{d}_h$, $h$ is any security having degree $\tilde{d}_h$, and if we define $q_{ij}^h=Q_{ih}Q_{jh}$, 
then the expected overlap $\langle o_{ij}^h\rangle_\Omega$ between institutions $i$ and $j$ restricted to securities having degree $\tilde{d}_h$ follows the binomial distribution
\begin{equation}
\pi_h(x|\tilde{n}_h,q_{ij}^h)=\binom{\tilde{n}_h}{x}[q_{ij}^h]^x[1-q_{ij}^h]^{\tilde{n}_h-x}.
\end{equation}
The overall distribution $\pi(\cdot|d_i,d_j)$ can now be more easily obtained as the sum of (much fewer than $S$) binomial random variables~\cite{DSBRV}:
if $\pi_{\le h}(\cdot|d_i,d_j)$ is the distribution of the overlap restricted to securities with degree smaller or equal than $h$, we have 
\begin{equation}
\pi_{\le h}(x|d_i,d_j)=\sum_{k=0}^x \pi_{\le 
h-1}(x-k|d_i,d_j)\pi_h(x|\tilde{n}_h,q_{ij}^h)
\label{pval_sum}
\end{equation}
and $\pi(\cdot|d_i,d_j)=\pi_{\le d_s^{max}}(\cdot|d_i,d_j)$. For this computation, it is useful to recall the peculiar recurrence relation of the binomial distribution: 
starting from $\pi_h(0|\tilde{n}_h,q_{ij}^h)=[1-q_{ij}^h]^{\tilde{n}_h}$, each subsequent probability is obtained through:
\begin{equation}
\pi_h(x|\tilde{n}_h,q_{ij}^h)=\frac{\tilde{n}_h-x+1}{x}\frac{q_{ij}^h}{1-q_{ij}^h}\pi_h(x-1|\tilde{n}_h,q_{ij}^h).
\end{equation}
Once the distribution $\pi(\cdot|d_i,d_j)$ is obtained, the p-value $P(o_{ij})$ can be associate to the overlap $o_{ij}$ using Eq.~(\ref{eq.pval}), 
and the corresponding link can be placed on the validated monopartite network provided that $P(o_{ij})\le P^*$. 
Note that since this computation is made on the whole network, {\em i.e.}, considering all the securities, we have a fairly large spectrum of possible p-values. 
Thus, also if we still use a threshold depending on the number of hypothesis tested (which however now scales just as $I^2$), 
we have a much higher resolution than in~\cite{Tumminello2011PLoS}, and can obtain non-empty and denser validated networks.

\section*{Acknowledgments}
We thank Fabio Saracco, Tiziano Squartini and Fr\'ed\'eric Abergel for useful discussions. 
D. Challet and S. Gualdi thank Luciano Pietronero together with the (ISC)-CNR team for their kind hospitality. 
S. Gualdi acknowledges support of Labex Louis Bachelier (project number ANR 11-LABX-0019). 
G. Cimini and R. Di Clemente acknowledge support from project GROWTHCOM (FP7-ICT, grant n. 611272). 
G. Cimini acknowledges support from projects MULTIPLEX (FP7-ICT, grant n. 317532) and DOLFINS (H2020-EU.1.2.2., grant n. 640772). 
The funders had no role in study design, data collection and analysis, decision to publish, or preparation of the manuscript. 

\section*{Author contributions statement}
S.G., G.C. R.D.C. and D.C. conceived the experiment. S.G., G.C., K.P. and R.D.C. 
conducted the experiments. S.G., G.C. and D.C. analysed the results and wrote 
the manuscript. 
All authors reviewed the manuscript.

\end{document}